\documentclass[a4paper,11pt]{article}
\usepackage{jcappub} 
\usepackage{lineno}
\usepackage{amsmath}
\usepackage{amsfonts}
\usepackage[english]{babel}
\usepackage{amssymb}
\usepackage{subcaption}
\usepackage{multicol}
\usepackage{float}
\usepackage[dvipsnames]{xcolor}
\usepackage{wrapfig}
\usepackage{graphicx}
\usepackage{listings}
\usepackage{multirow}
\usepackage[mathscr]{euscript}

\graphicspath{{Imagenes/}}

\hypersetup{
    colorlinks=true,
    linkcolor=red, 
    citecolor=blue, 
    urlcolor=purple 
}

\title{Comparing Hemispheres: Anisotropy in the deceleration parameter $q_0$}
\author[a]{Mauricio Lopez-Hernandez}

\author[a,b]{Josue De-Santiago}
    \affiliation[a]{ Departamento de F\'isica, Centro de Investigaci\'on y de Estudios Avanzados del I.P.N. 
                    Apartado Postal 14-740, 07000, Ciudad de M\'exico, M\'exico}
    \affiliation[b]{Secretar\'ia de Ciencia, Humanidades, Tecnolog\'ia e Innovaci\'on,
    			Av. Insurgentes Sur 1582, 03940, Ciudad de M\'exico, M\'exico}
\date{\today}

\emailAdd{mauricio.lopez@cinvestav.mx}
\emailAdd{Josue.desantiago@cinvestav.mx}

\abstract{We present a hemispherical comparison analysis of the deceleration parameter $q_0$ using the Pantheon+ sample of Type Ia supernovae to test the isotropy of cosmic acceleration and the robustness of redshift corrections. We detect significant directional variations in $q_0$ across all redshift frames. Even in the $z_{\mathrm{HD}}$ frame, where corrections for the CMB dipole and peculiar velocities are applied, a residual dipolar anisotropy persists with $\Delta q_0 = 0.112$ and a maximum signal to noise $S/N = 2.155$, aligned with the CMB dipole direction and decreasing with increasing minimum redshift cut. As expected, the anisotropy is stronger in the $z_{\mathrm{hel}}$ and $z_{\mathrm{CMB}}$ frames, where these kinematic corrections are incomplete, while the transition to $z_{\mathrm{HD}}$ substantially reduces the signal but does not fully remove it. We further show that inferring the dipole directly from the supernovae data yields $v_{\odot} = 307.26^{+32.00}_{-22.28},\mathrm{km\,s^{-1}}$ toward $(\mathrm{RA},\mathrm{DEC}) = (156.40^{+4.72}_{-4.71},, -3.38^{+5.54}_{-8.23})^\circ$, mildly discrepant with the Planck CMB dipole at the $\sim 1.9\sigma$ level and providing a better fit to the data. When this SNe inferred dipole is consistently incorporated into the redshift correction pipeline, the previously observed hemispherical anisotropy is largely suppressed, with the dipolar pattern disappearing and the maximum signal to noise reduced to $S/N \lesssim 1.75$, while the remaining fluctuations become patchy and consistent with statistical noise, suggesting that part of the original signal arises from residual mismatches in the modeling of the local velocity field rather than from a genuine large scale cosmological anisotropy. Since current redshift corrections rely on peculiar velocity reconstructions based on the density field and improved methodologies implemented in Pantheon+, our results suggest the presence of a coherent residual bulk flow not fully captured by these models. The alignment of the preferred directions with previously reported cosmic dipoles supports a kinematic interpretation, and highlights a previously neglected source of systematic uncertainty in low redshift supernova cosmology, with direct implications for the inference of cosmic anisotropies.}

\begin{document}
\maketitle
\flushbottom

\section{\label{sec:intro}Introduction}

The cosmological principle (CP), the assumption that the Universe is statistically homogeneous and isotropic on sufficiently large scales, has long served as a cornerstone of modern cosmology. When combined with General Relativity, it leads uniquely to the accepted concordance $\Lambda$CDM model through the Friedmann--Lema\^{\i}tre--Robertson--Walker (FLRW) metric. Within this framework, the large scale expansion of the Universe is fully characterized by time dependent, but direction and position independent, background quantities such as the Hubble parameter $H(z)$ and the deceleration parameter $q(z)$. Any statistically significant directional dependence in these observables would therefore signal either a breakdown of the cosmological principle or the presence of unaccounted for physical or observational effects. 

The present day deceleration parameter $q_0$ plays a central role in this context, as it directly characterizes the accelerated or decelerated nature of the cosmic expansion. Since the discovery of cosmic acceleration from Type Ia supernovae (SNe) observations \cite{Riess_1998, Perlmutter_1999}, $q_0$ has become a key cosmological parameter closely linked to the energy content of the Universe. Within the FLRW framework, $q_0$ is a global scalar quantity and is therefore expected to be isotropic when inferred from sufficiently large and well sampled data sets.

In recent years, however, the CP has been increasingly confronted by the growing volume and precision of cosmological observations, some of which appear to challenge its assumptions. Evidence for this has emerged from measurements in the cosmic microwave background (CMB) \cite{Gaztanaga_2024, PhysRevD.105.083508, PhysRevLett.93.221301, 10.1093/mnras/stv501, Planck_isotropy_refId0, Akrami_2014, PhysRevD.72.103002, PhysRevD.69.063516, Planck2018_isotropy_refId0, Mukherjee_2016}, SNe Ia \cite{MCCLURE2007533, PhysRevD.102.103525, Lopez-Hernandez_2025, OCOLGAIN2024101464, PhysRevD.88.083529, Dainotti_2021, Tsagas_Anisotropy_Pantheon+, Colin_refId0, 10.1093/mnras/stac1223, 10.1093/mnras/stad2788, PhysRevD.106.103527, HU_refId0, PhysRevD.107.023507, Eoin_hemisferios_Pantheon+,Schwarz_hemispheres,Eoin_hemisferios_Pantheon, Clocchiatti_2024}, the cosmic dipole anomaly \cite{Horstmann_2022,Sorrenti_2023, Singal_2022, Singal_2011, Bengaly_2018, Singal_PhysRevD.100.063501, Schwarz_refId0, Secrest_2021, Singa_10.1093/mnras/stac144, Secrest_2022, Colin_10.1093/mnras/stx1631, 9ygx-z2yq, Tiwari_2016, 10.1111/j.1365-2966.2012.22032.x}, structures that exceed $\Lambda$CDM expectations \cite{10.1093/mnras/stac2204, 10.1093/mnras/stv1421, 10.1093/mnras/staa2460, Laniakea_Tully, 10.1093/mnras/sts497, Pomarede_2020, Gott_III_2005, Horvath_refId0} and anomalous
bulk flows \cite{Kashlinsky_2008, Lavaux_2010, Kashlinsky_2010, Watkins_large_cosmic_flows} (for recent reviews on the validity of the cosmological principle see \cite{Kumar_Aluri_2023, PERIVOLAROPOULOS2022101659}). 

While such observations do not necessarily imply a fundamental violation of large scale isotropy, they highlight the need for careful interpretation. Directional signatures may arise from a combination of local structure, peculiar velocities, survey systematics, uneven sky coverage, or the choice of observational frame. In this broader context, tilted cosmological models \cite{Tsagas_2010, Tsagas_2011, Tsagas_2015, Tsagas_2021} provide a useful and physically motivated framework to explore how apparent anisotropies can emerge even when the underlying spacetime geometry remains close to FLRW. 

In tilted scenarios, the observer’s rest frame does not coincide with the cosmic rest frame, leading to relative bulk motions between different families of observers. Such relative motion effects alone can induce different locally inferred values of cosmological parameters in the two frames. In particular, observers embedded in large scale bulk flows may infer effective modifications to the local expansion history, including apparent acceleration even if the background Universe is globally decelerating. Although this effect is kinematic rather than due to a genuine dynamical acceleration, the relevant scales typically of order several hundred Mpc are sufficiently large to mimic a genuine cosmological signal.

A key prediction of this framework is that relative motions introduce Doppler anisotropies in locally inferred observables. Consequently, tilted observers are expected not only to measure directional variations in the Hubble expansion, but also to see an apparent dipole in the sky distribution of the deceleration parameter $q_0$ and the Hubble constant $H_0$, with an axis generically aligned with that of the CMB dipole if both effects share a common kinematic origin \cite{Tsagas_2022, 10.1093/mnras/stac922}. Indeed, such dipole like anisotropies have been reported in \cite{Eoin_hemisferios_Pantheon+, Eoin_hemisferios_Pantheon, Tsagas_Anisotropy_Pantheon+, Colin_refId0, PhysRevD.105.103510, Migkas_refId0, Boubel_2025, Rong-Gen_Cai_2012, zhao_aniso_q0}. 

Among the works cited above, several have explicitly targeted anisotropies in $q_0$ using complementary strategies and SNe compilations. Using the JLA sample, Ref. \cite{Colin_refId0} modeled $q_0$ as a monopole plus a redshift dependent dipole modulation, and found that the inferred dipole component dominates over any isotropic acceleration out to $z\sim 0.1$ and at $3.9\sigma$ of statistical    significance, with the preferred dipole direction lying within the CMB dipole. Ref. \cite{zhao_aniso_q0} instead constructed anisotropic $q_0$ maps from Union2 by splitting the sky into 12 Galactic regions and fitting $q_0$ in each subset; they reported a nonzero dipole amplitude $A_1 = 0.466^{+0.255}_{-0.205}$ at $\gtrsim 2\sigma$ but with a dipole direction that is nearly perpendicular to the CMB kinematic dipole, highlighting that inferred $q_0$ dipoles can depend sensitively on sky sampling and analysis choices. Similarly, using the Union2 compilation and the hemisphere comparison method, Ref. \cite{Rong-Gen_Cai_2012} quantified anisotropy through the normalized hemispherical difference $\Delta q_0/\bar{q}_0$ and identified a preferred axis at $\mathrm{RA} \simeq 196.5^\circ$, $\mathrm{DEC} \simeq -34^\circ$, with $\Delta q_0/\bar{q}_0 = 0.79^{+0.28}_{-0.27}$, finding that the signal is more pronounced at low redshift ($z \leq 0.2$) and largely insensitive to the assumed dark energy parametrization. More recently, when a dipolar component in $q_0$ is introduced in analyses of the Pantheon+ sample, \cite{Tsagas_Anisotropy_Pantheon+} report a strong statistical preference for such a term under likelihood ratio tests, with the direction sometimes closer to, but not necessarily fixed by, the CMB dipole, and with the inferred significance and direction depending on the chosen rest frame, redshift cuts, and light curve standardization assumptions. Taken together, these studies motivate treating $q_0$ anisotropy as an empirically testable signature whose amplitude, axis, and interpretation can be intertwined with redshift corrections, peculiar velocities modeling, and frame choices.

Motivated by these considerations, in this work we investigate the isotropy of the deceleration parameter $q_0$ using a hemispherical comparison approach applied to the Pantheon+ SNe Ia sample \cite{Pantheon+}. Building on earlier analyses of hemispherical anisotropies in the Hubble diagram and inspired by predictions of tilted cosmological models, we search for directional variations in $q_0$ and assess their statistical significance. Particular attention is paid to the role of redshift corrections, and peculiar velocity modeling. We introduce an extra dipolar velocity component to mitigate the inferred $q_0$ anisotropies.

The structure of this paper is as follows. Section \ref{sec:framework} introduces the analysis framework, including the Pantheon+ compilation and the procedure used to infer a preferred dipolar direction directly from the applied redshift corrections. Section \ref{sec:metodology} presents the methodology of the hemispherical analysis, describing the hemisphere comparison method, the visualization strategy, the role of minimum redshift cuts, and the implementation of the Pantheon+ covariance matrix in the dipole inference. Section \ref{sec:results} contains the main results, including the hemispherical anisotropy obtained in different redshift frames, the inferred SNe dipole and its impact on the anisotropy signal, and the residual bulk flow implied by these constraints. Finally, Section \ref{sec:conclusion} summarizes our findings and presents the conclusions.

\section{\label{sec:framework}Framework}

\subsection{\label{subsec:data}The Pantheon+ compilation}

In this work we employ the Pantheon+ SNe Ia compilation that consists of 1701 calibrated light curves corresponding to 1550 distinct SNe, spanning a wide redshift range from \(z \simeq 0.001\) to \(z \simeq 2.26\) \cite{Pantheon+}. 

The Pantheon+ analysis reevaluates and homogenizes all supernova redshifts \cite{Carr_redshifts}. It starts from redshifts measured in the reference frame of the solar system ($z_{\rm{hel}}$) and subtracts the velocity of the cosmic microwave background (CMB), yielding $z_{\rm{CMB}}$, the redshift measured by an observer at rest with respect to the CMB. The analysis then applies corrections for host galaxy peculiar velocities using an improved velocity field reconstruction based on the 2M++ density field \cite{Carrick_2M++}. This approach models both local flows and large scale bulk motions, providing realistic corrections even beyond the very low redshifts. This leads to the  final redshift used for cosmological analyses, denoted $z_{\mathrm{HD}}$, which aims to isolate the redshift component due solely to the cosmological expansion, effectively placing the host galaxies in their local cosmic rest frame and minimizing contamination from peculiar motions.

These improvements are particularly relevant for analyses sensitive to directional effects, since residual dipolar signatures from uncorrected peculiar motions could otherwise mimic or obscure genuine cosmological anisotropies \cite{10.1093/mnras/stac1223,Carr_redshifts,Rameez_EPJS_Sne}.

\begin{figure*}
    \centering
    \begin{subfigure}[t]{0.57\textwidth}
        \centering
        \includegraphics[width=\linewidth]{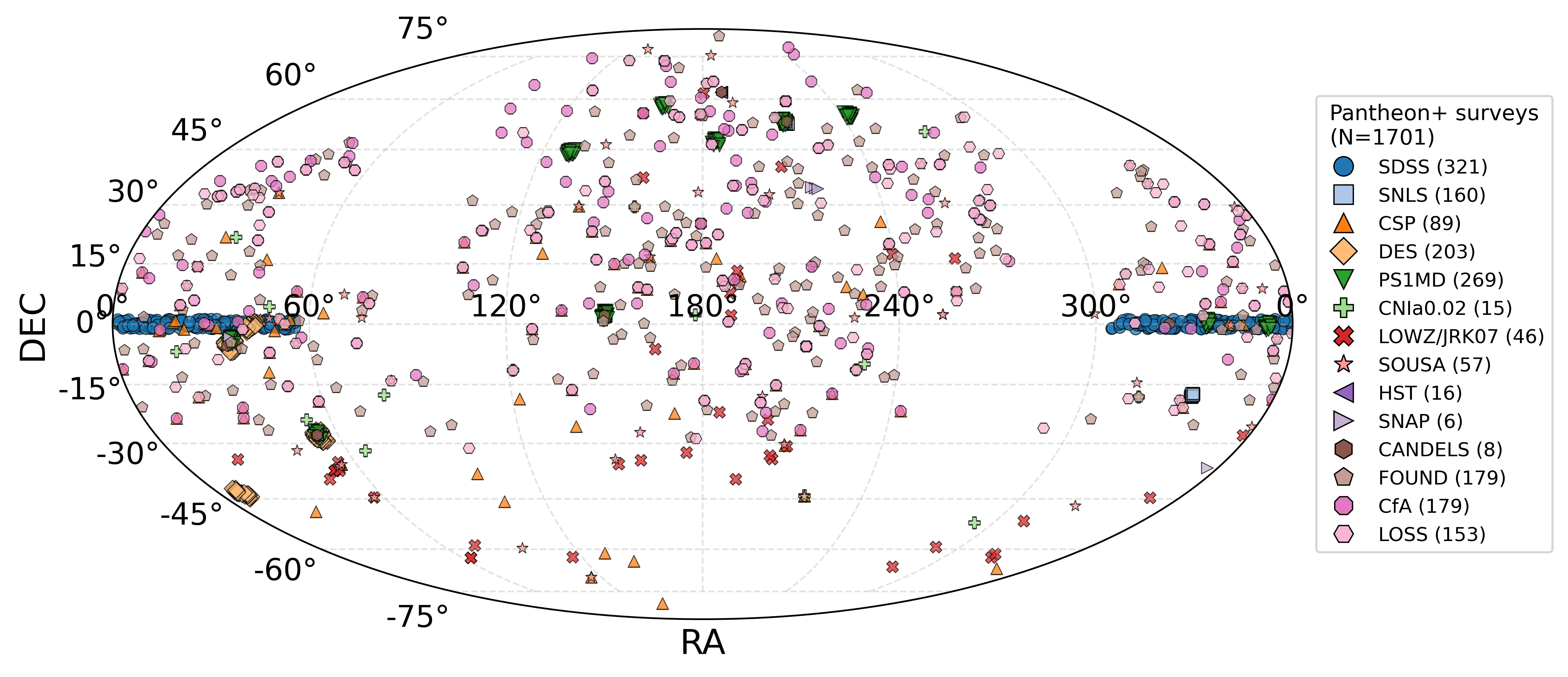}
    \end{subfigure}
    \hfill
    \begin{subfigure}[t]{0.42\textwidth}
        \centering
        \includegraphics[width=\linewidth]{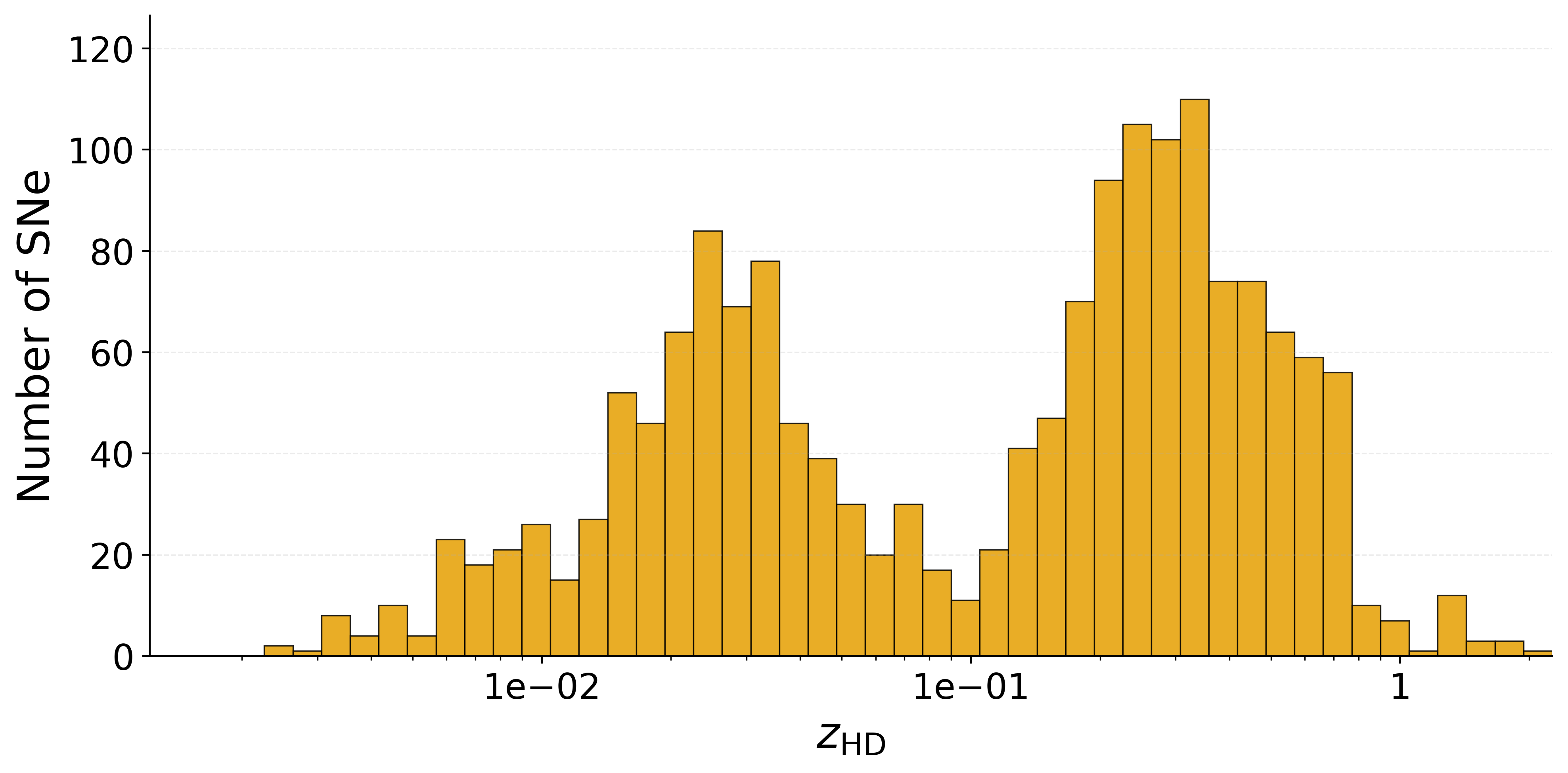}
    \end{subfigure}
    \caption{
    Left panel: sky distribution of the Pantheon+ SNe Ia in equatorial coordinates, shown in a Mollweide projection and symbols color coded by survey.
    Right panel: redshift distribution of the Pantheon+ sample.
    }
    \label{fig:pantheon_sky_redshifts}
\end{figure*}

Pantheon+ exhibits a non uniform distribution both in redshift and on the sky. As shown in the right panel of Fig.~\ref{fig:pantheon_sky_redshifts}, the redshift distribution is strongly weighted toward low and intermediate redshifts, with a substantial fraction of the SNe located at $z \lesssim 0.1$, and a progressively sparser population toward higher redshifts. This behavior reflects the heterogeneous origin of the compilation, which combines targeted low redshift surveys with deeper, wide field programs optimized for cosmological measurements. As a result, while the sample spans more than three orders of magnitude in redshift, its statistical constraining power is dominated by low and intermediate redshifts.

On the other hand, the angular distribution is likewise markedly inhomogeneous; see the left panel of Fig.~\ref{fig:pantheon_sky_redshifts}. SNe at very low redshift ($z \lesssim 0.05$) are distributed over a wide range of directions, reflecting the contribution of multiple targeted nearby surveys. In contrast, a large fraction of the intermediate redshift sample ($0.05 \lesssim z \lesssim 0.35$) is concentrated along the celestial equator, tracing the footprint of the SDSS observing strategy. At higher redshifts, the sky coverage becomes increasingly patchy and anisotropic, with SNe clustered in a few localized deep fields. As a result, extended sky regions remain poorly sampled at high redshift.

These features imply that different directions on the sky probe distinct redshift ranges and survey combinations. Consequently, any analysis sensitive to directional variations in cosmological parameters must carefully account for the coupled angular and redshift inhomogeneities of the data. In particular, apparent anisotropic signals could arise from the interplay between survey geometry, redshift dependent uncertainties, and local velocity effects, as well as from genuine departures from isotropy. 

To infer a value of $q_0$ from the Pantheon+ data, we model the observed corrected apparent magnitudes, $m_\textrm{B}^{\mathrm{obs}}(z)$, using a cosmographic expansion of the luminosity distance including terms up to third order in redshift. Following the formulation adopted by the SH0ES collaboration \cite{Riess_2022}, the theoretical apparent magnitude is expressed as
\begin{equation}
m_\textrm{B}^{\mathrm{th}}(z) = 5\log_{10}\!\left[c\,z\left(1 + \frac{1}{2}(1-q_0)z - \frac{1}{6}\left(1 - q_0 - 3q_0^2 + j_0\right)z^2\right)\right] + \mathcal{M},
\label{eq:mB_cosmogra}
\end{equation}
where $c$ denotes the speed of light and $j_0$ is the present day jerk parameter. The parameter
\begin{equation}
\mathcal{M} \equiv -5\log_{10}H_0 + \textrm{M}_\textrm{B} + 25
\label{eq:M_parameter}
\end{equation}
absorbs the dependence on $H_0$ and the absolute magnitude $M_B$ into a single nuisance parameter, following standard practice in local cosmographic analyses.

\subsection{\label{subsec:SNe_dipole}Inferring the SNe dipole from redshift corrections}

In this subsection we present an internal consistency test of the redshift correction pipeline by allowing the dipole that best maps the observed heliocentric redshifts of the Pantheon+ SNe into an isotropic Hubble diagram frame to be inferred directly from the data, rather than being fixed to the values measured from the CMB temperature dipole by Planck \cite{Planck_2018_I}. We refer to this quantity as the \emph{SNe dipole}, to distinguish it from the standard CMB dipole. In general, these two need not be identical. In the ideal case where the local velocity field is perfectly modeled and all anisotropic low redshift contributions are fully captured by the peculiar velocity corrections, the inferred SNe dipole would be expected to coincide with the standard CMB dipole. In practice, however, while the CMB dipole is conventionally interpreted as tracing the motion of the Solar System with respect to the CMB rest frame, the dipole preferred by the SNe sample may also absorb residual anisotropic contributions present in the low redshift velocity field, such as coherent bulk flows or limitations in the reconstruction of the peculiar velocity field. 

The peculiar velocity corrections adopted in Pantheon+ rely on a reconstructed velocity field that, while state of the art, is subject to several known limitations \cite{Carr_redshifts}. First, the reconstruction is not fully model independent, since mapping galaxy redshifts into distances and converting the observed galaxy density field into a peculiar velocity field requires assumptions about the fiducial background cosmology, the growth rate of structure, galaxy bias, and the validity of the linear theory relation between density and velocity. Second, the density field is only directly mapped within a finite survey volume, but the region where peculiar velocities are needed extends beyond the domain where the mass distribution is observationally constrained. As a result, contributions from structures outside the reconstructed volume must be approximated through an effective external bulk flow and an assumed radial decay consistent with $\Lambda$CDM, introducing additional modelling uncertainty. Third, the sampling of the galaxy distribution is incomplete and spatially non uniform, leading to regions where the inferred density and velocity fields are less reliable. Finally, the reconstruction is based largely on linear theory and smoothed fields, which may not fully capture nonlinear motions and environmental effects on small scales. Consequently, although these corrections substantially improve the treatment of nearby SNe, residual anisotropic signatures may remain in the corrected redshifts.

The motivation for this analysis is therefore twofold. First, it provides a direct test of whether the Pantheon+ SNe data are statistically consistent with using the standard CMB dipole. Second, it allows us to construct an alternative set of Hubble diagram redshifts based on the dipole preferred by the SNe themselves, which can then be used to reassess the hemispherical results for the deceleration parameter $q_0$. A significant discrepancy between the inferred SNe dipole and the CMB dipole would indicate that the effective frame selected by the SNe data differs from the standard one, potentially due to residual large scale motions not fully captured by the adopted PV model.

Following the Pantheon+ analysis (see \cite{Carr_redshifts}), we correct the observed heliocentric redshifts $z_{\mathrm{hel}}$ for the motion of the Solar System with respect to an hypothetical reference frame.
In the standard pipeline, this motion is taken to be the one inferred from the CMB dipole. Here, instead, we allow the data to determine the effective dipole preferred by the SNe sample, which we interpret as the preferred rest frame. Its amplitude $v_{\odot}$ and direction $\mathrm{RA}_{\rm dip}, \mathrm{DEC}_{\rm dip}$ are treated as free parameters.

For a SN located along the line of sight $\hat{\mathbf{n}}_{\rm SN}$, the redshift is modified  due to
the projection of the dipole velocity onto that direction,
\begin{equation}
\beta_{\odot} \equiv \frac{v_{\odot}}{c}\,
\hat{\mathbf{n}}_{\rm SN}\!\cdot\!\hat{\mathbf{n}}_{\rm dip},
\label{eq:beta_sol}
\end{equation}
where $\hat{\mathbf{n}}_{\rm dip}$ is the unit vector pointing toward the apex of the SNe dipole, specified by right ascension $\mathrm{RA}_{\rm dip}$ and declination $\mathrm{DEC}_{\rm dip}$. The sign convention is such that $\beta_{\odot}>0$ corresponds to motion toward the source. Since the correction depends on the line of sight through the scalar product above, each SN receives a different redshift shift according to its angular position relative to the inferred dipole direction.

The corresponding kinematic redshift factor relating the heliocentric frame to the preferred SNe frame is
\begin{equation}
1+z_{\odot} \;=\;
\sqrt{\frac{1-\beta_{\odot}}{1+\beta_{\odot}}},
\label{eq:z_sol}
\end{equation}
where we use the exact relativistic expression.

The redshift corrected for the observer's motion, denoted $z_{\mathrm{SN}}$, analogous to $z_{\mathrm{CMB}}$, follows from the multiplicative combination of redshift factors,
\begin{equation}
(1+z_{\mathrm{hel}}) \;=\; (1+z_{\mathrm{SN}})(1+z_{\odot}),
\qquad\Rightarrow\qquad
z_{\mathrm{SN}} \;=\; \frac{1+z_{\mathrm{hel}}}{1+z_{\odot}}-1.
\label{eq:zhel_to_zsn}
\end{equation}

Once the SNe frame redshifts are specified, we correct for the peculiar velocities of the host galaxies to obtain the Hubble diagram redshifts $z_{\mathrm{HD}}$. For this purpose, we recompute the reconstructed peculiar velocity field derived from the 2M++ density field \cite{Carrick_2M++}\footnote{The peculiar velocity method developed in Pantheon+ is available at \url{https://github.com/KSaid-1/pvhub}.} at every parameter evaluation. This is necessary because the predicted peculiar velocity for each SN is evaluated at that object's observed sky position and redshift. Consequently, the inferred peculiar velocities change whenever the dipole parameters are varied.

The model returns peculiar velocity along the line of sight $v_p$ for each host galaxy, then this velocity is converted into a peculiar redshift using the low velocity approximation $z_p \;\simeq\; v_p/c$, which is sufficient for the present analysis, since peculiar velocities are typically of the order of a few hundred $\mathrm{km\,s^{-1}}$.

Finally, we compute the Hubble diagram redshifts by removing the peculiar contribution multiplicatively,
\begin{equation}
z_{\mathrm{HD}} \;=\; \frac{1+z_{\mathrm{SN}}}{1+z_p}-1.
\label{eq:zsn_to_zhd}
\end{equation}

We restrict the analysis to SNe satisfying $0.01<z_{\mathrm{HD}}<0.8$, ensuring that the inference is insensitive to extreme local velocity effects at very low redshift. Because both the transformation from $z_{\mathrm{hel}}$ to $z_{\mathrm{SN}}$ and the subsequent evaluation of $v_p$ depend on the dipole parameters, any variation in the inferred SNe dipole propagates self consistently through the full redshift correction pipeline.

\section{\label{sec:metodology}Metodology}
\subsection{\label{subsec:hemisphere_method}Hemisphere comparison method}

To search for possible directional variations in the cosmic deceleration parameter, $q_0$, we adopt the hemisphere comparison method, following earlier works that applied similar approaches to tests of cosmological isotropy using SNe data (see, e.g., \cite{Eoin_hemisferios_Pantheon+,Schwarz_hemispheres,Eoin_hemisferios_Pantheon,I_Antoniou_2010, Clocchiatti_2024}). 

We construct a grid of trial directions on the sky by uniformly sampling right ascension (RA) and declination (DEC) in equatorial coordinates. 
We divided RA into 30 equally spaced values over the range $[0^\circ,360^\circ)$, while we sampled DEC with 15 evenly spaced values between $-90^\circ$ and $90^\circ$. Each grid point defines a unit vector $\hat{\mathbf{n}}_j$. Likewise, each SNe in Pantheon+ is associated with a unit vector $\hat{\mathbf{r}}_i$ constructed from its (RA, DEC) coordinates. For a given grid direction $\hat{\mathbf{n}}_j$, we divided the SNe sample into two hemispheres according to the sign of the scalar product $\hat{\mathbf{n}}_j \cdot\hat{\mathbf{r}}_i$. SNe satisfying $\hat{\mathbf{n}}_j \cdot\hat{\mathbf{r}}_i \geq 0$ are assigned to the hemisphere centered on $\hat{\mathbf{n}}_j$, while those with $\hat{\mathbf{n}}_j \cdot\hat{\mathbf{r}}_i < 0$ to the complementary hemisphere. We repeated this procedure independently for each grid direction, yielding two subsets of SNe per direction.

For each hemisphere, we model the observed corrected apparent magnitudes using Eq.~\eqref{eq:mB_cosmogra}, fixing the jerk parameter to $j_0=1$ as expected in a $\Lambda$CDM cosmology. Fixing $j_0$ provides a stable parametrization of the expansion history while allowing the analysis to focus on variations in $q_0$. Moreover, since the same value of $j_0$ is imposed for all hemispheres, this assumption does not introduce any directional dependence and therefore does not bias the search for anisotropies in the inferred values of $q_0$. We note, however, that recent constraints from dynamical dark energy parametrizations \cite{rodrigues2025cosmography, DESI_DR2} suggest values of $j_0$ significantly different from unity. In addition, $j_0$ and $q_0$ are intrinsically correlated in cosmographic parametrizations. This indicates that variations in $j_0$ can be partially absorbed by shifts in $q_0$, potentially affecting its inferred value. Therefore, fixing $j_0=1$ not only simplifies the parametrization but also removes this degeneracy, stabilizing the inference of $q_0$ across hemispheres. 

Consistent with the SH0ES analysis, we restrict to redshifts with $z \leq 0.8$, ensuring the validity of the truncated cosmographic expansion and limiting the impact of higher order terms. Over this range, SH0ES reports a measurement of $q_0 = -0.51 \pm 0.024$ \cite{Riess_2022}, which provides a useful benchmark for our hemispherical fits.

For each hemisphere defined by a given sky direction, we estimate the parameters $(q_0,\mathcal{M})$ using a Markov Chain Monte Carlo (MCMC) analysis restricted to the corresponding subset of SNe in the hemisphere. We defined the $\chi^2$ function as
\begin{equation}
\chi^2 = \Delta\mathbf{m}^{\mathrm{T}}\,\mathbf{C}^{-1}\,\Delta\mathbf{m},
\label{eq:chi2_hemisphere}
\end{equation}
where $\Delta\mathbf{m}= m_\textrm{B}^{\mathrm{obs}}(z) - m_\textrm{B}^{\mathrm{th}}(z)$ denotes the vector of residuals between observed and model magnitudes, and $\mathbf{C}$ is the covariance matrix including both statistical and systematic uncertainties. For each hemisphere, we obtained $\mathbf{C}$ by selecting the rows and columns of the full Pantheon+ covariance matrix corresponding to the SNe belonging to that hemisphere. This procedure effectively treats the covariance matrix as if only those SNe had been observed, neglecting correlations with SNe located in the opposite hemisphere.

Posterior distributions are sampled using the MCMC algorithm implemented in the Python package \texttt{emcee} \cite{emcee_Foreman-Mackey_2013}. Flat priors are assumed on both parameters, with broad bounds chosen to encompass all physically relevant values. For each hemisphere, parameter estimates and uncertainties are obtained from the marginalized posterior distributions, where the central values correspond to the median and the uncertainties are given by the 16th and 84th percentiles. Repeating this procedure over all grid directions yields a sky map of median values of the deceleration parameter, enabling a direct assessment of hemispherical asymmetries and preferred directions in the inferred $q_0$.

\subsection{Visualization}

To visualize the angular dependence of the hemispherical asymmetry in the deceleration parameter, we construct maps over the full sky derived from the results. For each trial direction $\hat{\mathbf{n}}_j$ on the sky, we define the antisymmetric quantity
\begin{equation}
\Delta q_0(\hat{\mathbf{n}}_j) \equiv q_0(\hat{\mathbf{n}}_j) - q_0(-\hat{\mathbf{n}}_j),
\label{eq:Delta_q0}
\end{equation}
where $q_0(\hat{\mathbf{n}}_j)$ and $q_0(-\hat{\mathbf{n}}_j)$ denote the median $q_0$ values obtained from the two complementary hemispheres associated with the direction $\hat{\mathbf{n}}_j$ and its antipode, respectively. By construction, this quantity satisfies $\Delta q_0(-\hat{\mathbf{n}}_j) = -\Delta q_0(\hat{\mathbf{n}}_j)$.

The discrete set of $\Delta q_0$ values obtained on the sky grid is interpolated onto a regular grid in equatorial coordinates and displayed using a Mollweide projection. This representation provides an intuitive view of the spatial structure of the inferred anisotropy and allows for the identification of directions that maximize $\Delta q_0$.

In addition, we construct a complementary map that quantifies the statistical significance of the asymmetry. For each independent sky direction we define a signal to noise ratio (S/N) as 
\begin{equation}
S/N(\hat{\mathbf{n}}_j) \equiv 
\frac{|q_0(\hat{\mathbf{n}}_j) - q_0(-\hat{\mathbf{n}}_j)|}
{\sqrt{\sigma_{q_0}^2(\hat{\mathbf{n}}_j) + \sigma_{q_0}^2(-\hat{\mathbf{n}}_j)}},
\label{eq:sigma_q0}
\end{equation}
where $\sigma_{q_0}(\hat{\mathbf{n}}_j)$ and $\sigma_{q_0}(-\hat{\mathbf{n}}_j)$ are the marginalized uncertainties on $q_0$ inferred from each hemisphere. By construction, this quantity is symmetric under the transformation $\hat{\mathbf{n}}_j \rightarrow -\hat{\mathbf{n}}_j$ and always positive.

Both maps are displayed using Mollweide projections in equatorial coordinates, with the same sky grid and interpolation scheme.

\begin{figure}[t]
    \centering
    \includegraphics[width=0.7\linewidth]{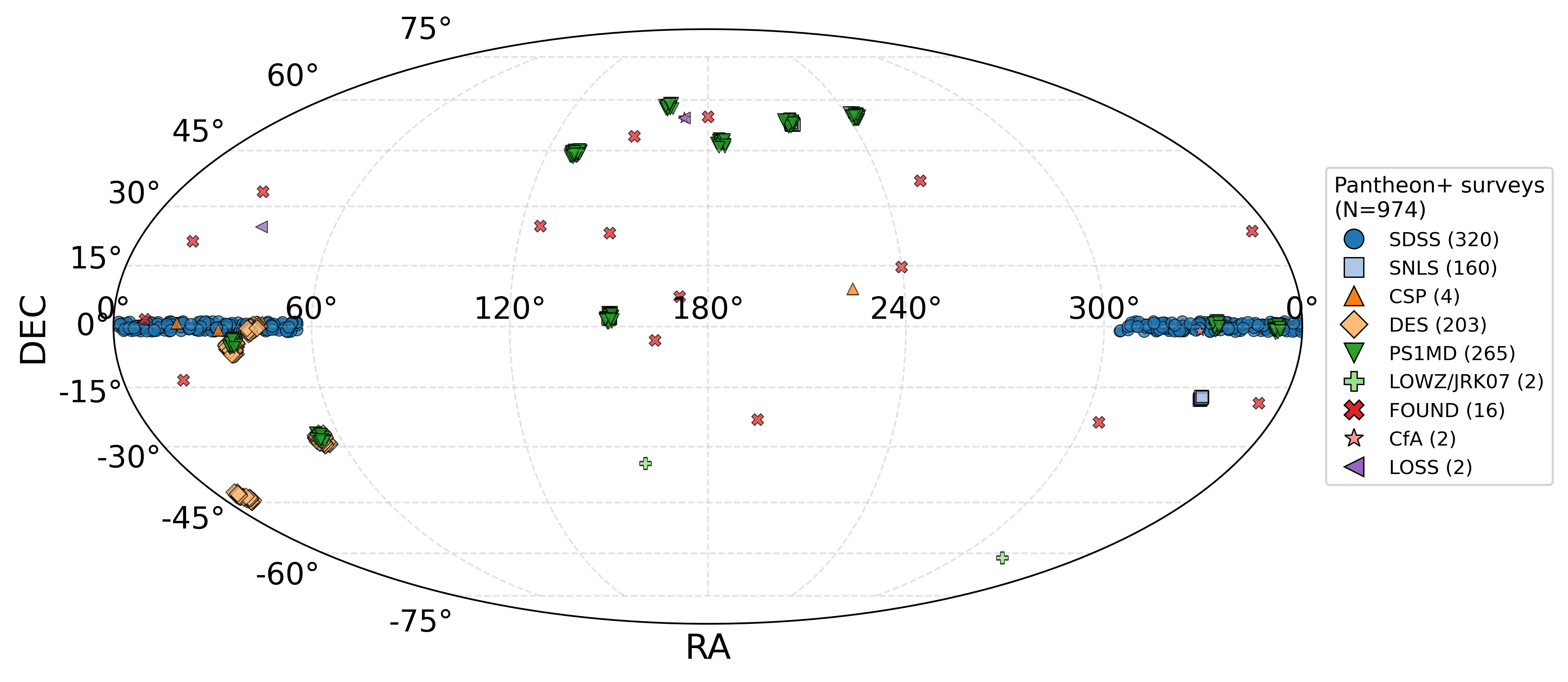}
    \caption{
    Sky distribution of Pantheon+ SNe Ia in equatorial coordinates for the redshift range
    $0.07 < z \leq 0.8$. Different symbols indicate the contributing surveys. The figure illustrates the highly non uniform angular coverage at higher redshifts, with the sample dominated by a limited number of survey footprints and extended regions of the sky devoid of SNe.
    }
    \label{fig:sky_zmin_0.07}
\end{figure}

\subsection{Minimum redshift cuts}

A critical aspect of any analysis based on SNe Ia is the choice of the minimum redshift included in the sample, as peculiar velocities can introduce significant scatter and potential biases at very low redshifts. Pantheon+ includes SNe with $z_{\mathrm{HD}} \geq 0.01$, while explicitly correcting for peculiar velocities using a reconstructed velocity field based on the 2M++ density map \cite{Pantheon+, Carr_redshifts, Carrick_2M++}. Although Pantheon+ demonstrates that cosmological parameter estimates remain robust under this treatment, it is well established that the regime $z_{\mathrm{HD}} \lesssim 0.01$ is particularly sensitive to residual velocity effects (see Fig.~4 in \cite{Pantheon+}). 

In the context of the distance ladder employed by the SH0ES collaboration to measure $H_0$, the nearby Hubble diagram at $z_{\rm  HD} \lesssim 0.01$ is not used directly, since the peculiar velocities dominate over the Hubble flow in this regime \cite{Riess_2022}. Instead, SH0ES adopts a conservative cut at $z_{\mathrm{HD}} > 0.023$ to define the Hubble flow sample, thereby minimizing sensitivity to local velocity fields without relying strongly on their explicit modeling.

Motivated by these considerations, we explore the stability of the hemispherical anisotropy signal under a series of increasingly conservative lower redshift cuts. In all cases, we fix the maximum redshift to $z_{\mathrm{HD}}^{\mathrm{max}} = 0.8$, consistent with the validity range of the truncated cosmographic expansion used in SH0ES. The redshift interval considered therefore takes the form
\begin{equation}
z_{\mathrm{HD}}^{\mathrm{min}} < z \leq 0.8,
\end{equation}
with $z_{\mathrm{HD}}^{\mathrm{min}}$ varied across the values $z_{\mathrm{HD}}^{\mathrm{min}} = 0.01,\; 0.023,\; 0.03,\; 0.05,\; \text{and } 0.07$.

The lowest cut, $z_{\mathrm{HD}}^{\mathrm{min}} = 0.01$, corresponds to the effective minimum redshift used in Pantheon+, while $z_{\mathrm{HD}}^{\mathrm{min}} = 0.023$ matches the SH0ES definition of the Hubble flow sample. The additional cuts at higher redshift are introduced to test whether any observed dipolar pattern in $q_0$ weakens or disappears as progressively more local SNe are excluded.

We choose the upper limit $z_{\mathrm{HD}}^{\mathrm{min}} = 0.07$ as a practical maximum for this exercise. As illustrated in Fig.~\ref{fig:sky_zmin_0.07}, SNe in the range $0.07 < z \leq 0.8$ no longer provide a quasi uniform angular coverage of the sky. Instead, the sample becomes increasingly dominated by a limited number of survey footprints and localized patches. Adopting a more restrictive higher $z_{\mathrm{HD}}^{\mathrm{min}}$ would therefore severely degrade the angular completeness of the data and compromise the interpretability of the hemispherical comparisons. This behavior is consistent with the heterogeneous survey composition of Pantheon+.

\subsection{Pantheon+ Covariance matrix in the inference of the SNe dipole} 

For the inference of the SNe dipole parameters, discussed in subsection \ref{subsec:SNe_dipole}, we adopt the same covariance matrix as used in the standard Pantheon+ likelihood, namely the full statistical plus systematic covariance matrix. This choice ensures direct consistency with the baseline Pantheon+ treatment and allows for a meaningful internal comparison between the standard redshift correction pipeline and the generalized approach explored here.

The Pantheon+ covariance matrix includes several contributions associated with photometric calibration uncertainties, intrinsic scatter of the SN population, survey systematics. We note, however, that this matrix includes contributions associated with the modeling of peculiar velocities, which were constructed assuming fixed CMB dipole parameters as measured by Planck. In principle, a fully self consistent analysis would require recomputing the covariance matrix for each set of dipole parameters, as changes in the heliocentric to SNe   transformation and in the predicted peculiar velocities can propagate into the velocity related covariance terms. Such a recalculation is computationally nontrivial and lies beyond the scope of the present work.

The use of the fixed Pantheon+ covariance matrix therefore represents a controlled approximation. On one hand, it guarantees consistency with the published Pantheon+ results and avoids introducing additional modeling assumptions. On the other hand, it may in principle introduce a mild bias or underestimate uncertainties in the inferred SNe dipole parameters. Nevertheless, we expect this effect to be small, since the dominant contributions to the covariance are not strongly sensitive to modest variations in the dipole amplitude or direction, and because peculiar velocities uncertainties mainly affect the lowest redshift regime, which we mitigate by cutting the redshifts smaller than 0.01.

Within this context, this analysis should be interpreted as an internal consistency test of the Pantheon+ redshift correction framework, rather than as a fully self calibrated re estimation of the peculiar velocity covariance. 

\section{\label{sec:results}Results}
\subsection{\texorpdfstring{Hemispherical anisotropy in the $z_{\mathrm{HD}}$ frame}{Hemispherical anisotropy in the z HD frame}}

We examined the hemispherical anisotropy in $q_0$ within the $z_{\mathrm{HD}}$ reference frame, which, as recalled, provides the most appropriate basis for assessing whether residual directional signatures in $q_0$ persist after accounting for known kinematic effects.

\begin{figure*}
\centering

\begin{subfigure}{\textwidth}
    \centering
    \includegraphics[width=0.43\textwidth]{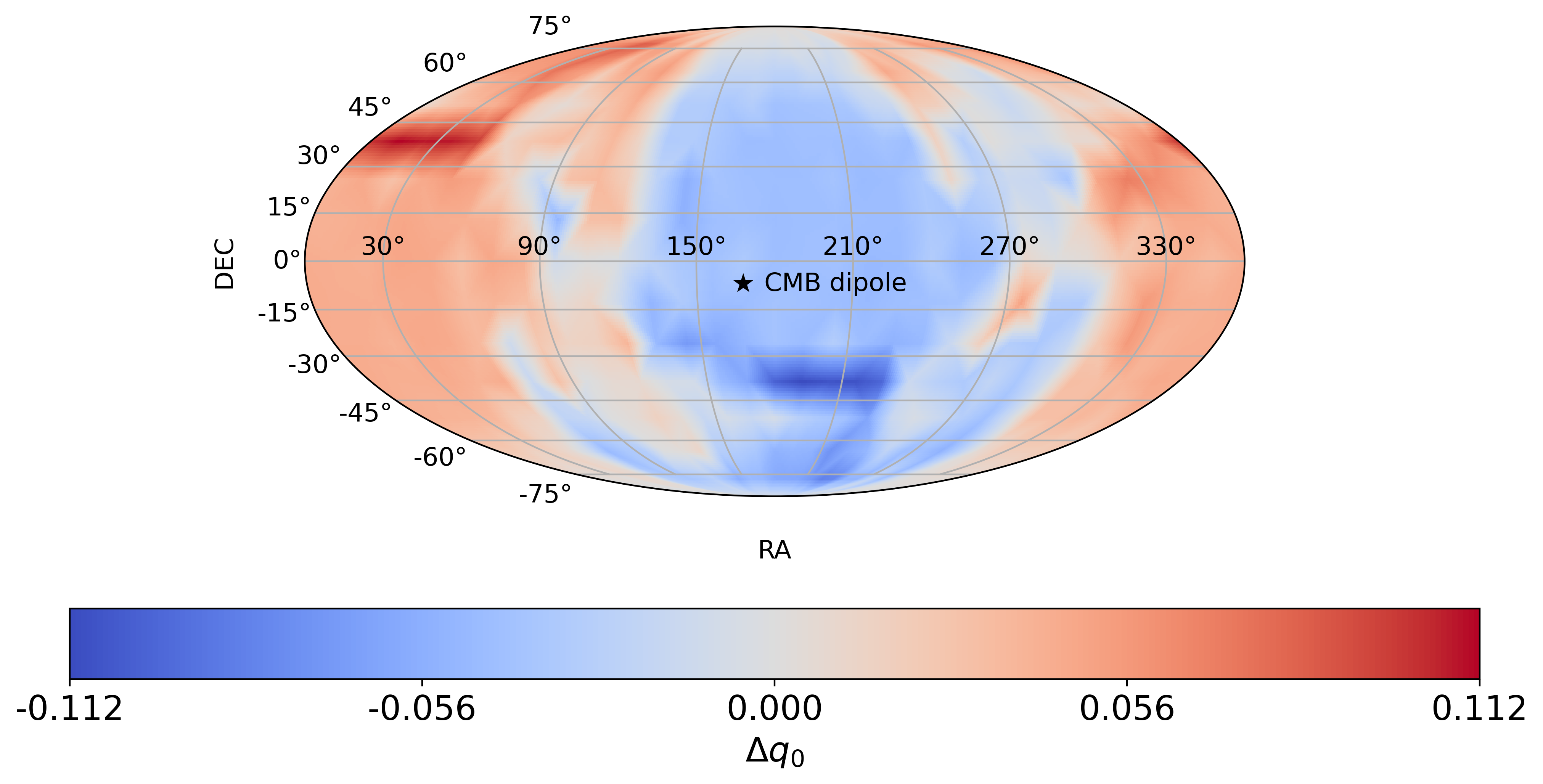}
    \hfill
    \includegraphics[width=0.43\textwidth]{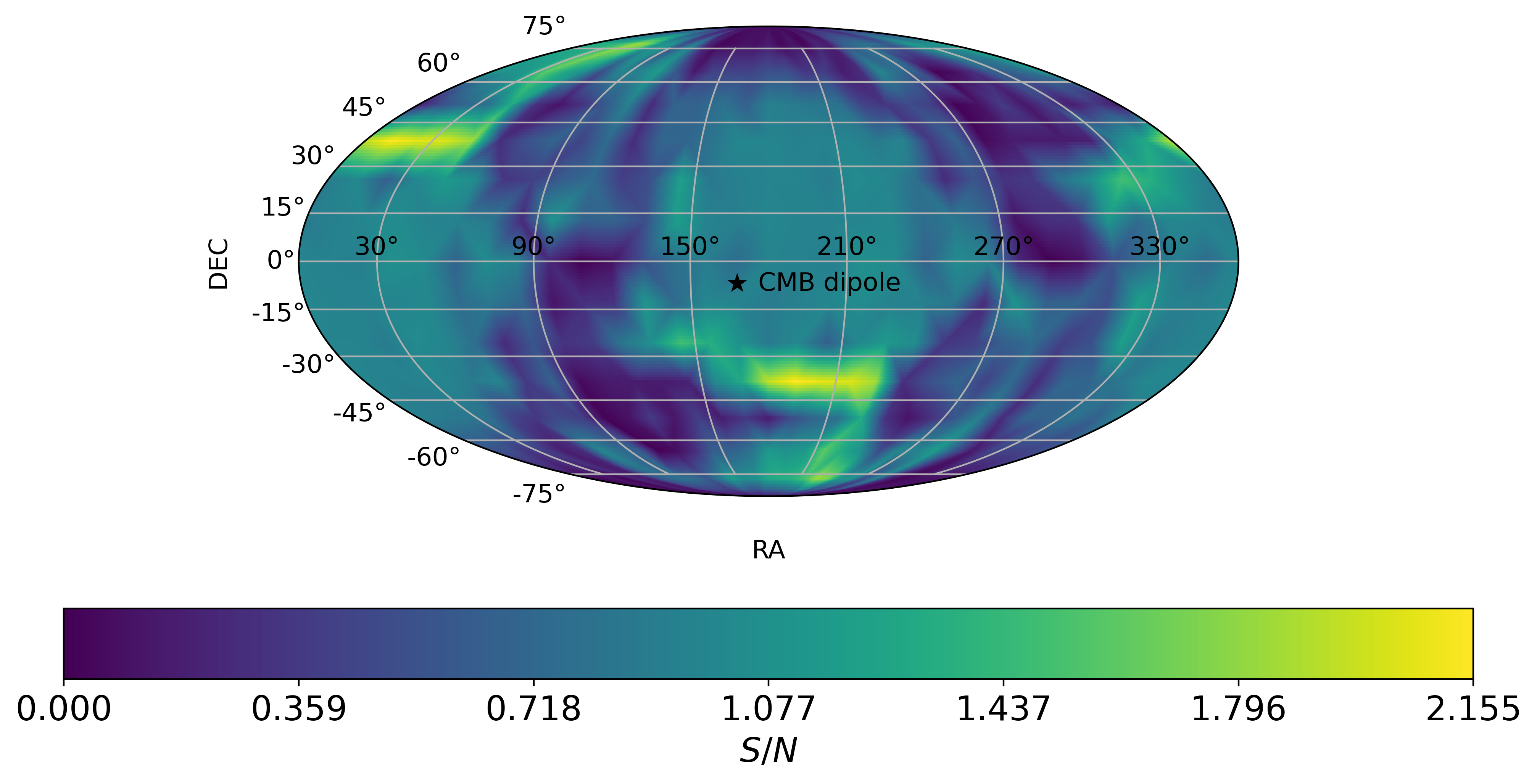}
    \caption{$0.01 < z_\textrm{HD} \leq 0.8$}
\end{subfigure}

\begin{subfigure}{\textwidth}
    \centering
    \includegraphics[width=0.43\textwidth]{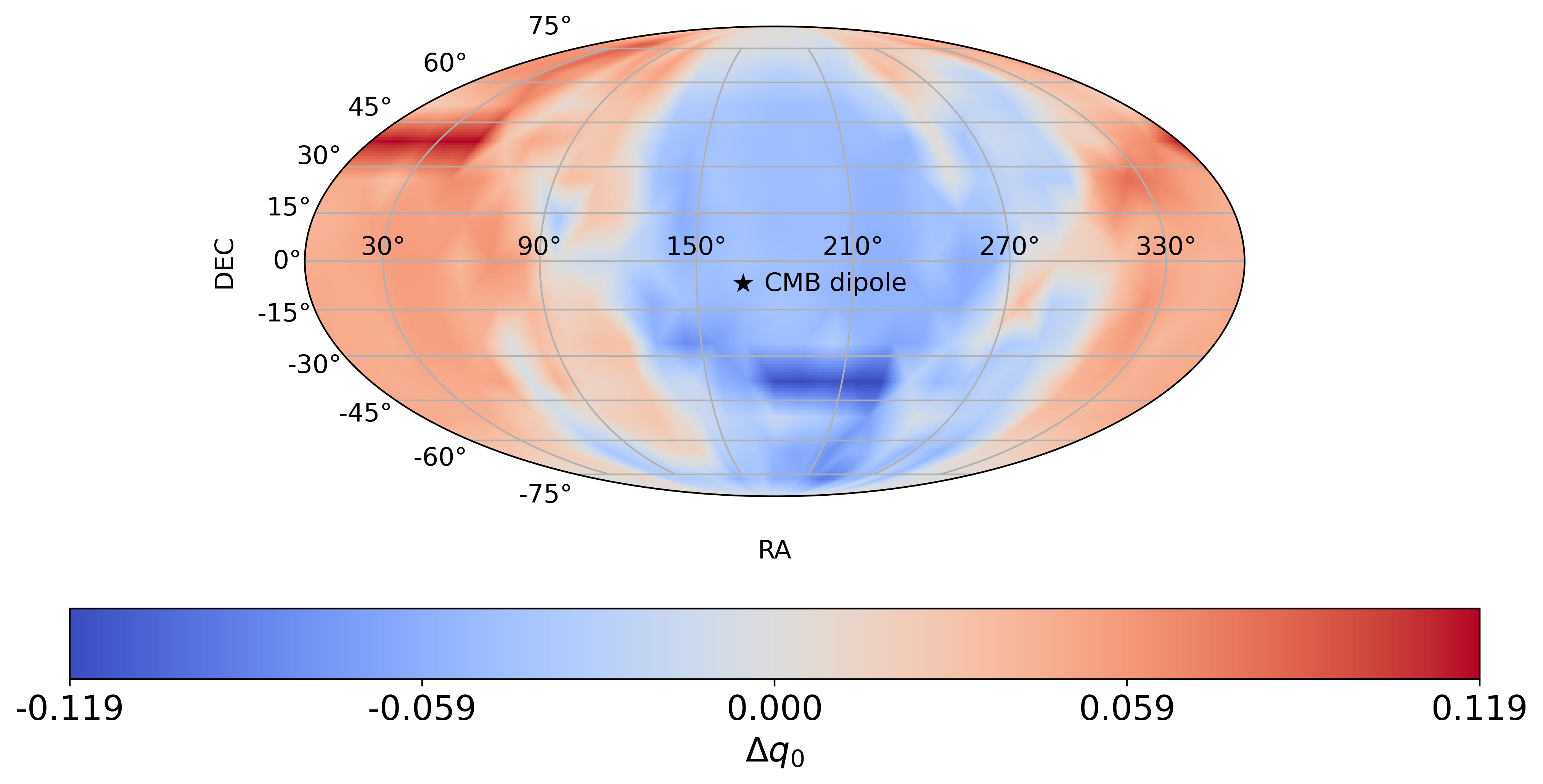}
    \hfill
    \includegraphics[width=0.43\textwidth]{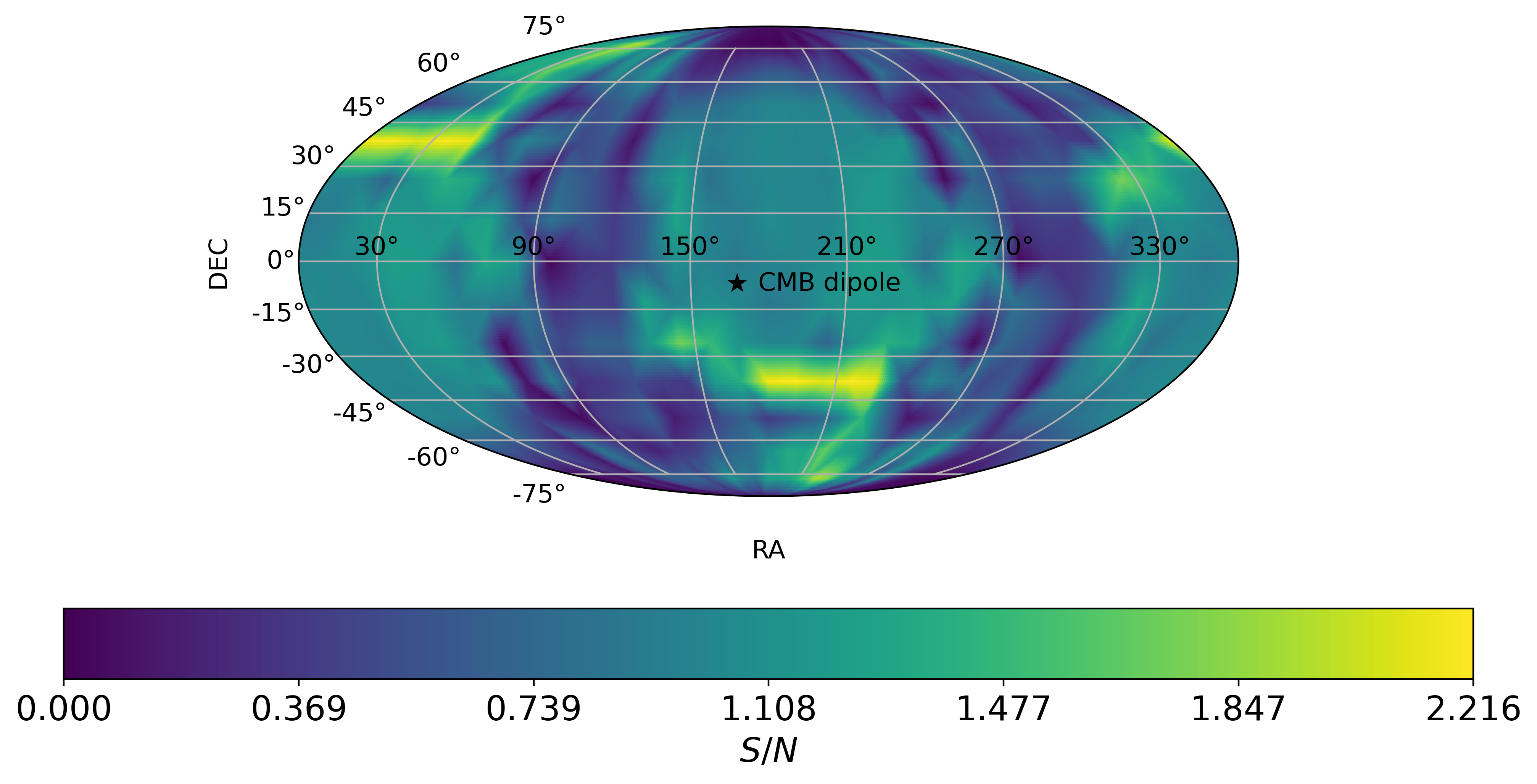}
    \caption{$0.023 < z_\textrm{HD} \leq 0.8$}
\end{subfigure}

\begin{subfigure}{\textwidth}
    \centering
    \includegraphics[width=0.43\textwidth]{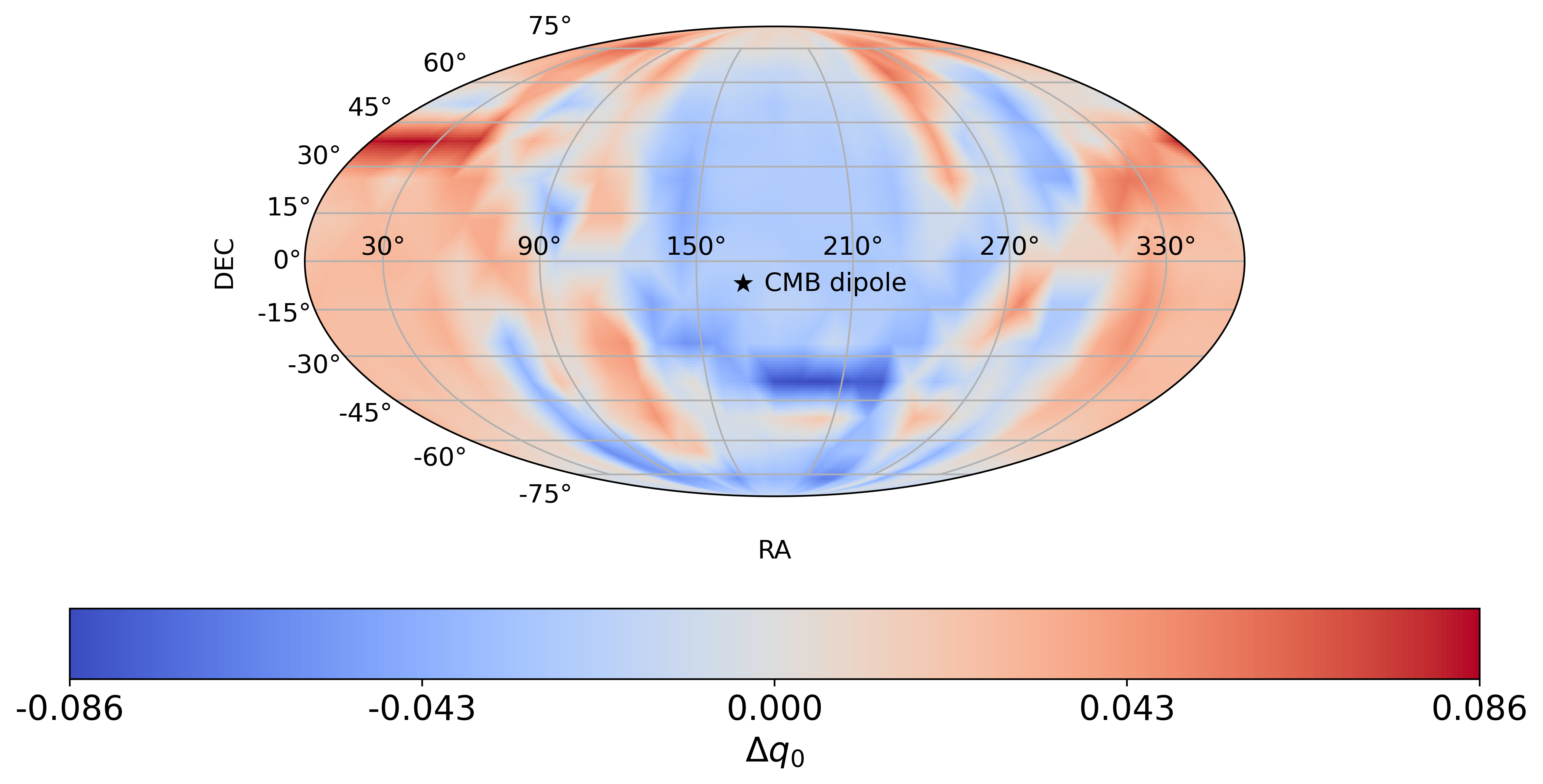}
    \hfill
    \includegraphics[width=0.43\textwidth]{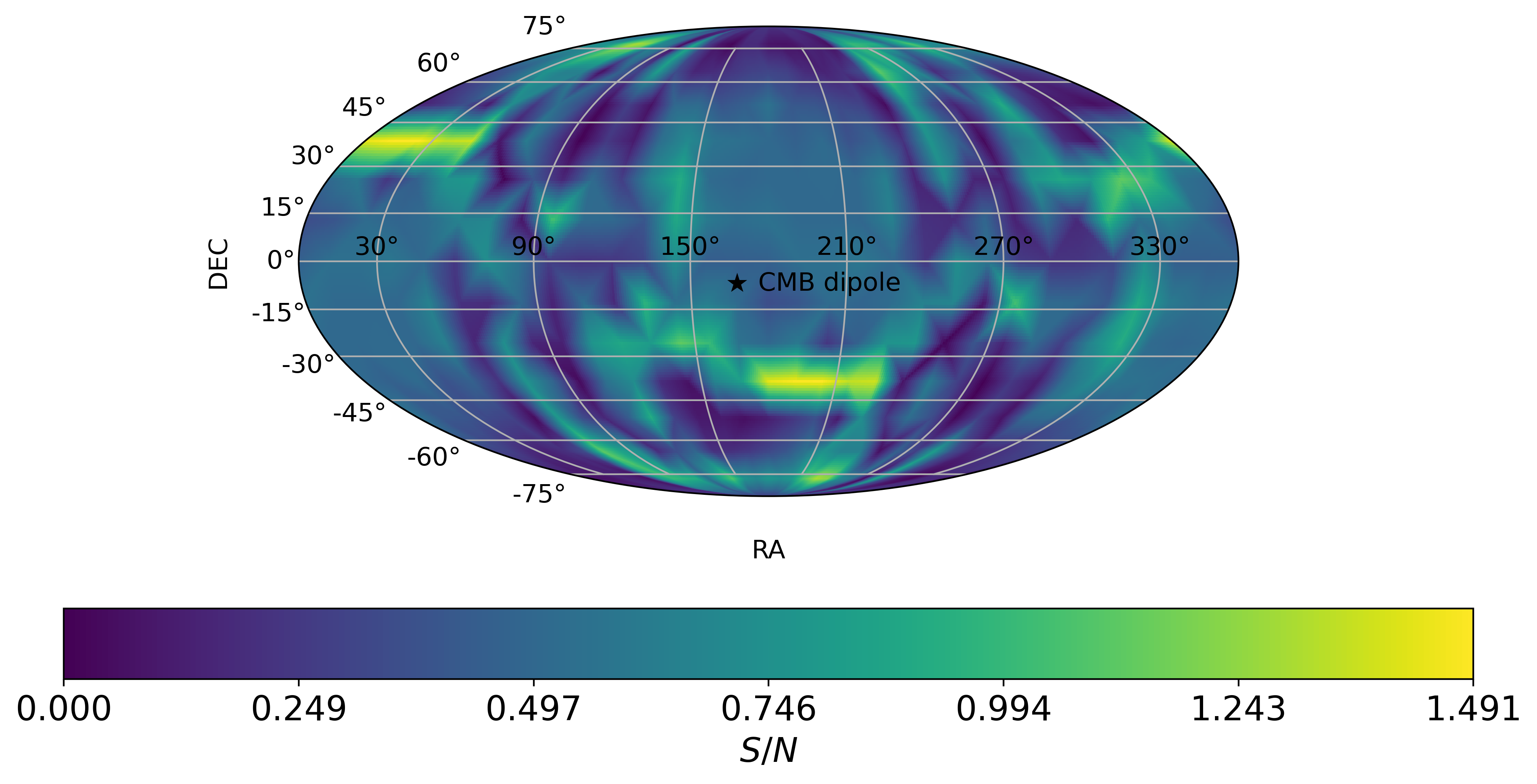}
    \caption{$0.03 < z_\textrm{HD} \leq 0.8$}
\end{subfigure}

\begin{subfigure}{\textwidth}
    \centering
    \includegraphics[width=0.43\textwidth]{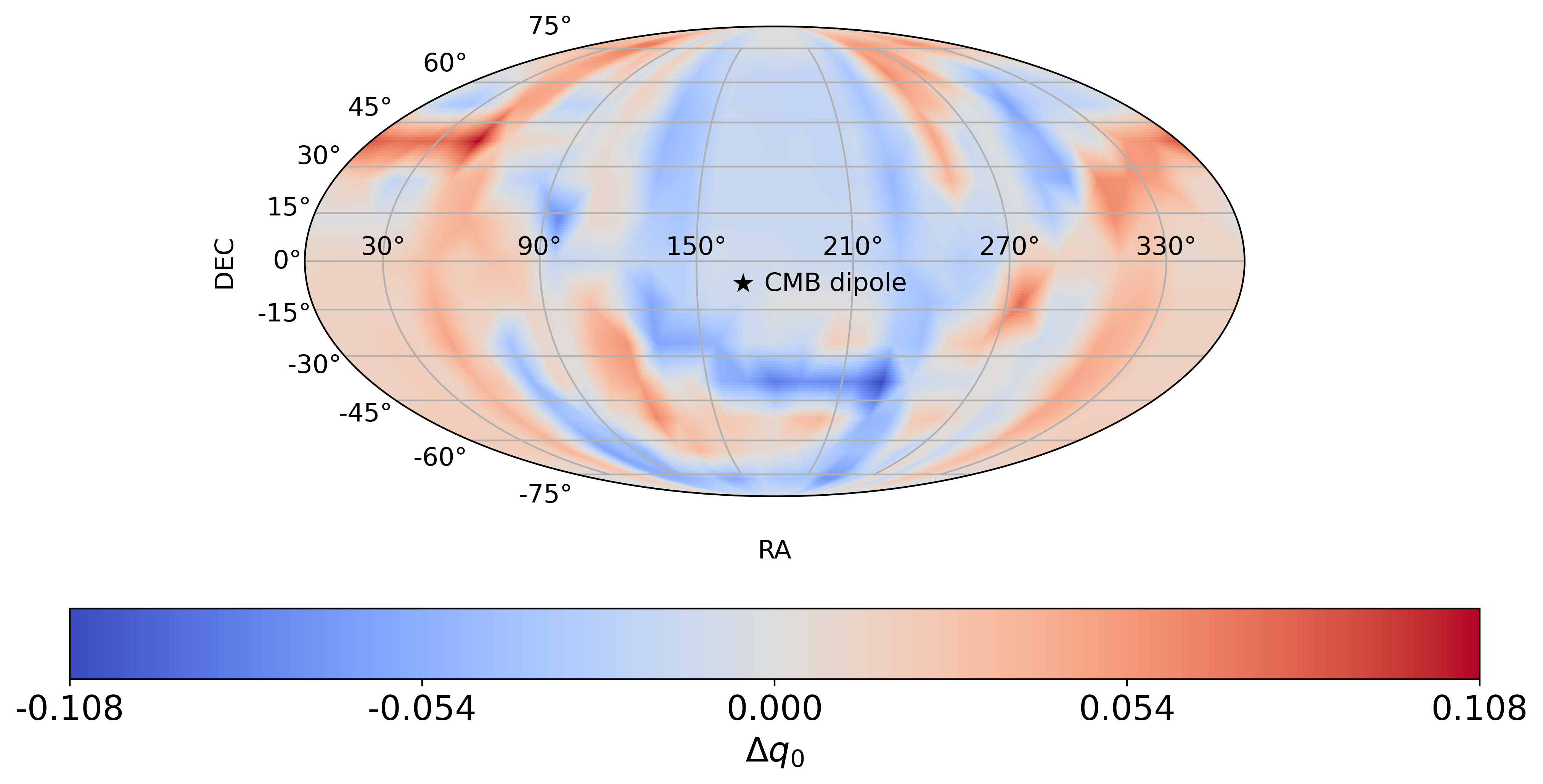}
    \hfill
    \includegraphics[width=0.43\textwidth]{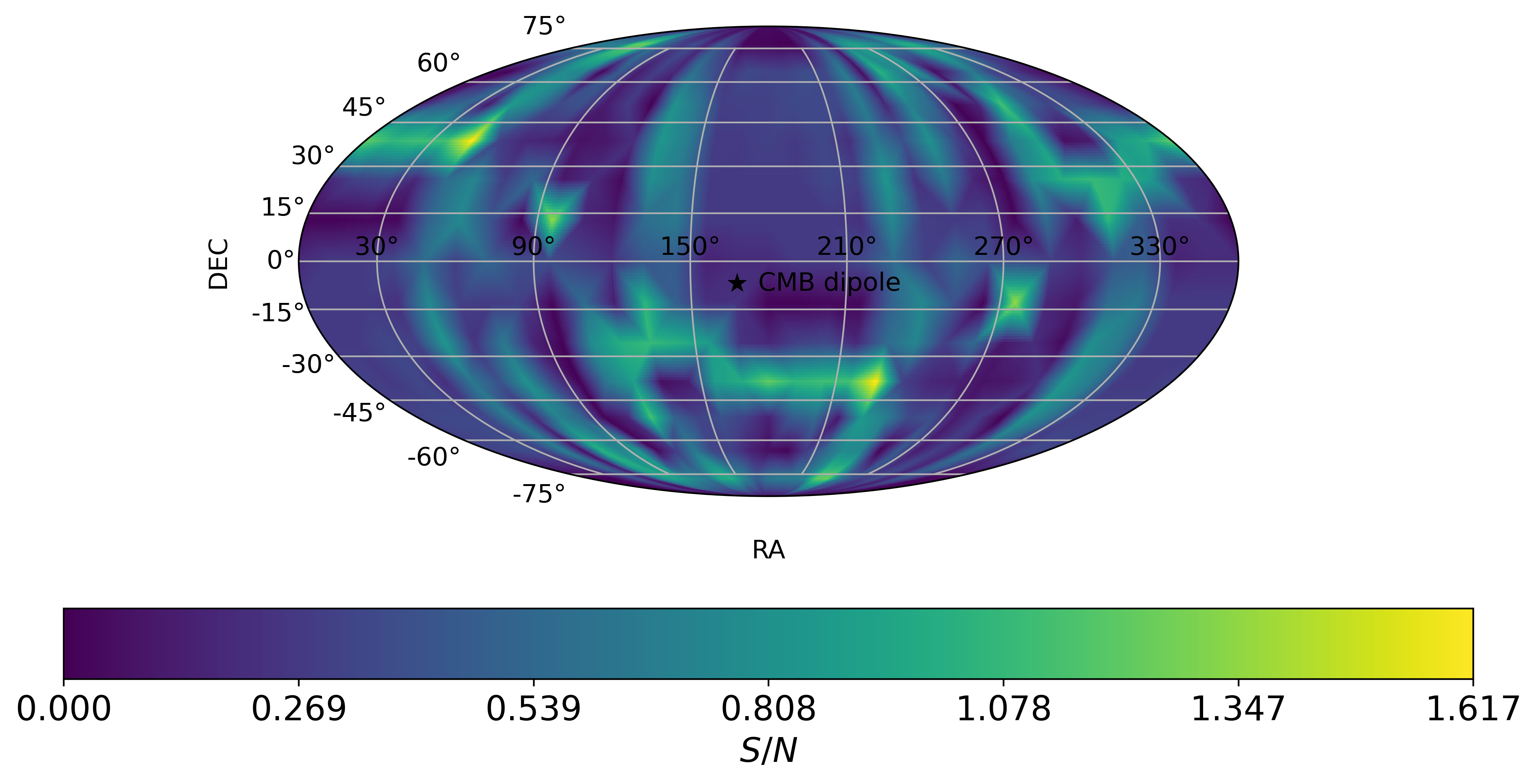}
    \caption{$0.05 < z_\textrm{HD} \leq 0.8$}
\end{subfigure}

\begin{subfigure}{\textwidth}
    \centering
    \includegraphics[width=0.43\textwidth]{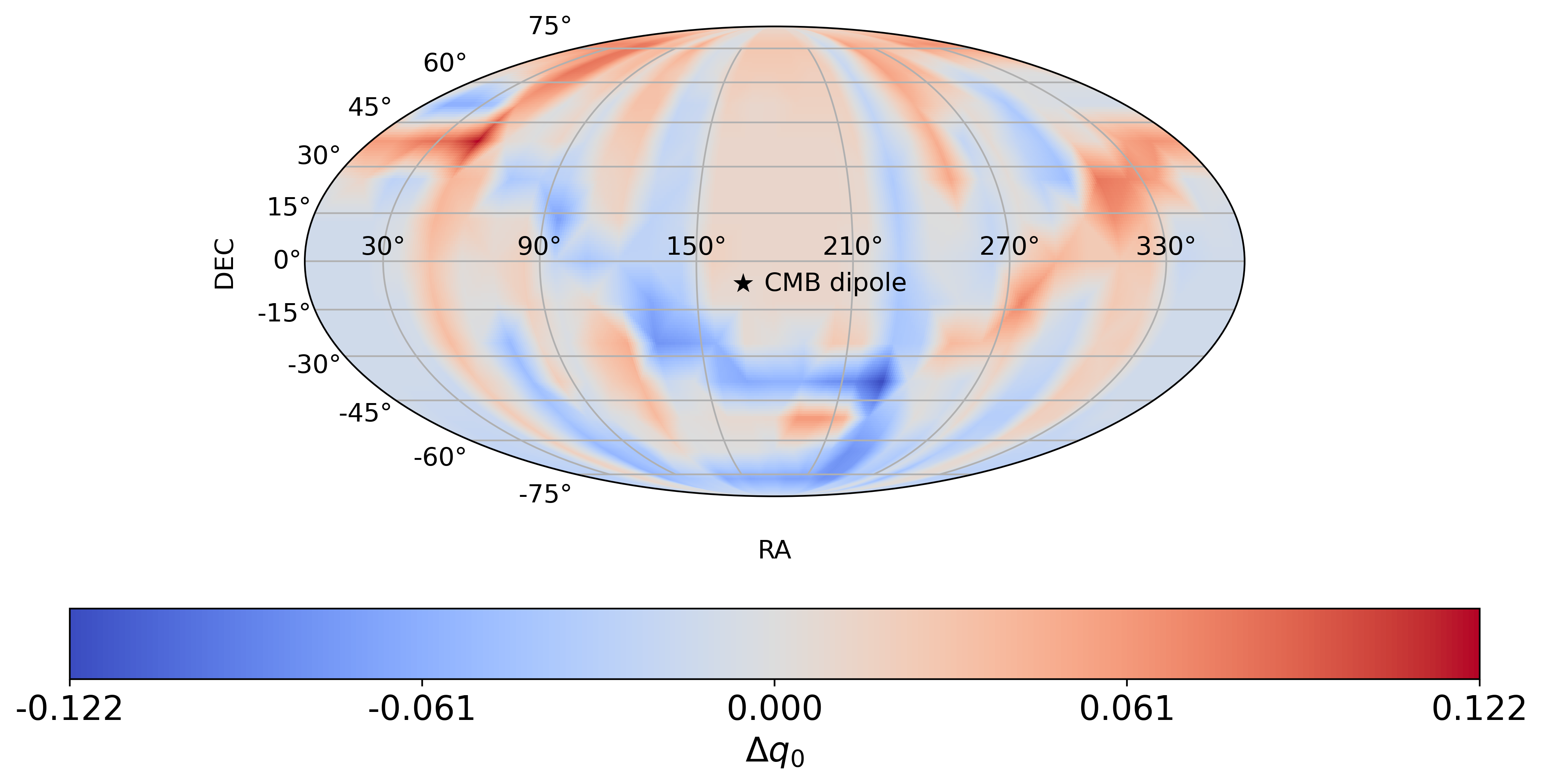}
    \hfill
    \includegraphics[width=0.43\textwidth]{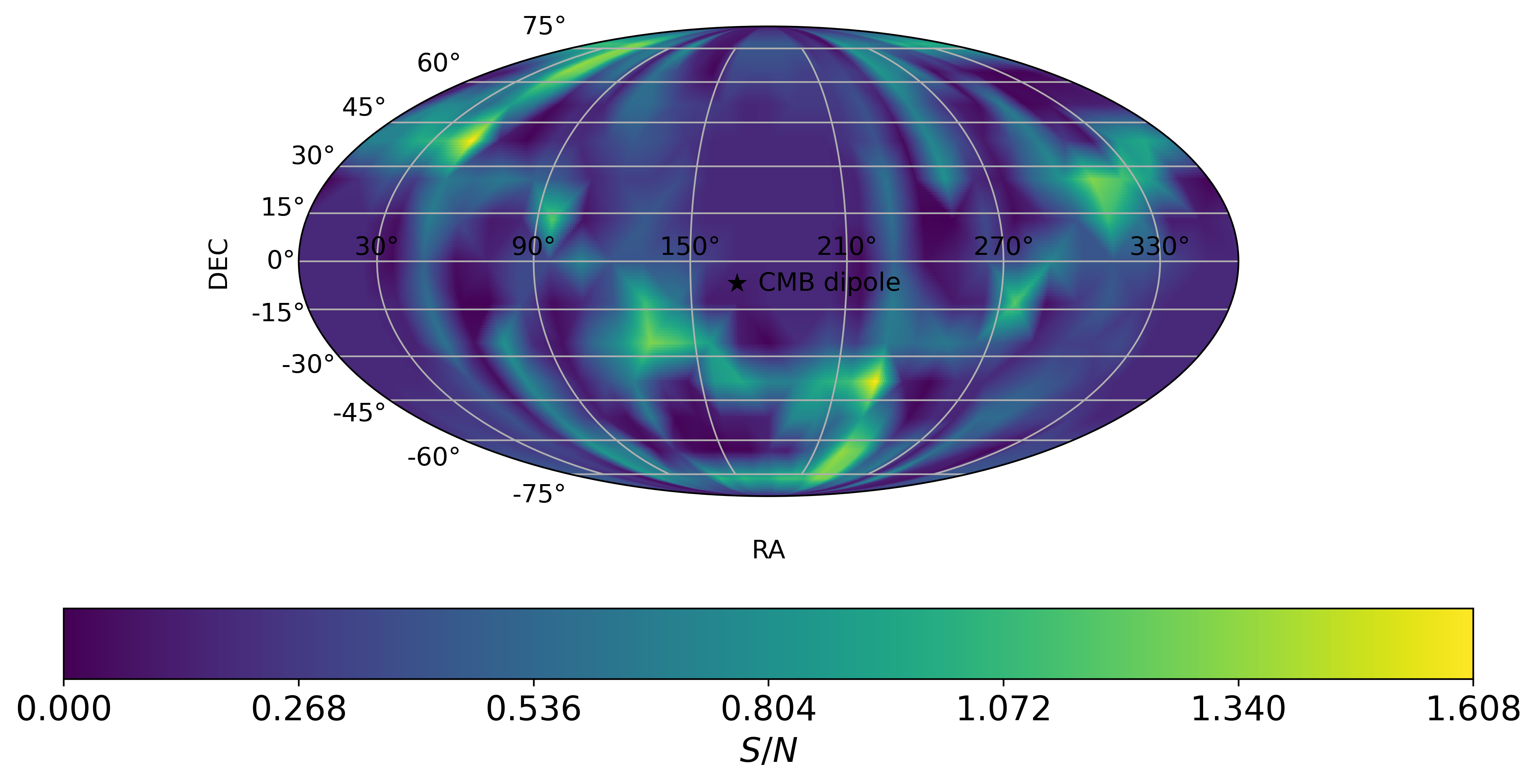}
    \caption{$0.07 < z_\textrm{HD} \leq 0.8$}
\end{subfigure}

\caption{
Sky maps of the hemispherical contrast in $q_0$.
Left column: $\Delta q_0 = q_0(\hat{\mathbf{n}}) - q_0(-\hat{\mathbf{n}})$.
Right column: local signal to noise ratio $S/N$.
Each row corresponds to a different minimum redshift cut $z_{\mathrm{HD}}^{\mathrm{min}}$, as indicated. The star marks the direction of the CMB dipole.
}
\label{fig:deltaq0_sigma_zcuts}
\end{figure*}

Fig.~\ref{fig:deltaq0_sigma_zcuts} shows sky maps of the hemispherical contrast $\Delta q_0$ and the corresponding local signal to noise ratio $S/N$ for a sequence of increasingly conservative values of $z_{\mathrm{HD}}^{\mathrm{min}}$. For the lowest cuts, $0.01 < z_{\mathrm{HD}} \leq 0.8$ and $0.023 < z_{\mathrm{HD}} \leq 0.8$, a clear dipolar pattern is visible in both maps. In these cases, hemispheres whose symmetry axis lies close to the direction of the CMB dipole tend to exhibit systematically lower values of $q_0$ than their antipodal counterparts, corresponding to negative values of $\Delta q_0$ in that region of the sky.

As we increased the minimum redshift cut to $z_{\mathrm{HD}}^{\mathrm{min}} = 0.03$, the hemispherical pattern becomes less pronounced, although a residual alignment with the CMB dipole direction can still be identified. For even higher cuts, $z_{\mathrm{HD}}^{\mathrm{min}} = 0.05$ and $0.07$, the dipolar structure progressively weakens and eventually becomes difficult to distinguish from statistical fluctuations. This behavior is qualitatively expected if the observed anisotropy is primarily driven by local or intermediate scale effects that diminish as we remove the nearby SNe.

\begin{table}[b]
\centering
\begin{tabular}{ccccc}
\hline\hline
$z_{\mathrm{HD}}^{\mathrm{min}}$ & $(\Delta q_0)_\textrm{min}$ & $(S/N)_\textrm{max}$ & RA [deg] & DEC [deg] \\
\hline
0.01  & $-0.112$ & $2.155$ & 192.18 & $-38.44$ \\
0.023 & $-0.119$ & $2.216$ & 216.54 & $-38.44$ \\
0.03  & $-0.086$ & $1.491$ & 203.91 & $-38.44$ \\
0.05  & $-0.108$ & $1.617$ & 227.37 & $-38.44$ \\
0.07  & $-0.122$ & $1.608$ & 227.37 & $-38.44$ \\
\hline\hline
\end{tabular}
\caption{Extrema values of the hemispherical contrast $\Delta q_0$ \eqref{eq:Delta_q0} and local signal to noise ratio $S/N$ \eqref{eq:sigma_q0} in the $z_{\mathrm{HD}}$ frame for different minimum redshift cuts. Quoted directions correspond to the hemisphere axis associated with the minimum $\Delta q_0$, i.e. the direction where
$q_0$ attains its most negative value. The opposite directions correspond to those that maximize $\Delta q_0$, or equivalently, where $q_0$ is least negative.}
\label{tab:deltaq0_sigma_extrema}
\end{table}

At the same time, the loss of a clear dipolar pattern at higher $z_{\mathrm{HD}}^{\mathrm{min}}$ must be interpreted with caution. As illustrated by the sky distribution shown in Fig.~\ref{fig:sky_zmin_0.07}, the angular coverage of the sample becomes increasingly incomplete and dominated by a small number of survey footprints. Such strong inhomogeneities reduce the statistical power of hemispherical comparisons and can naturally suppress an underlying dipolar pattern, even if a physical signal were present. 

It is worth noting that, for the lowest redshift cuts, the sign of the hemispherical contrast indicates that $q_0$ is more negative in the direction along the CMB dipole than in the opposite. This qualitative trend differs from the sign of the dipole amplitude reported in the parametric dipole fits of \cite{Tsagas_Anisotropy_Pantheon+}, although the two approaches are not directly equivalent in terms of methodology and underlying assumptions. 

The extrema of the hemispherical contrast and local signal to noise ratio, for each redshift cut, are summarized in Table~\ref{tab:deltaq0_sigma_extrema}. For all values of $z_{\mathrm{HD}}^{\mathrm{min}}$, the directions associated with the maximum signal to noise ratio $(S/N)_{\max}$ cluster around similar DEC values, with $\mathrm{DEC}\simeq -38.44^\circ$, while the RA varies modestly as the minimum redshift cut is increased. In our runs we find that the direction of $(S/N)_{\mathrm{max}}$ coincides either with the direction of $(\Delta q_0)_{\min}$ or with its antipode $(\Delta q_0)_{\max}$.\footnote{This behavior is expected because the signal to noise ratio depends on the absolute value $|\Delta q_0|$ and therefore cannot distinguish between a dipole axis and its opposite direction.} This behavior reflects the finite angular resolution of the discrete grid used to define the hemispherical axes, in which we sampled DEC at only 15 fixed values. Consequently, the repeated appearance of the same DEC should not be interpreted as a precise localization of a preferred physical direction, but rather as an indication that the underlying anisotropic signal is consistently captured by the same DEC band of the grid. The stability of the preferred DEC value across different redshift cuts primarily reflects the geometric persistence imposed by the discrete grid, even though the amplitude and the signal to noise ratio of the dipolar pattern decrease for more restrictive samples.

\subsection{\texorpdfstring{$z_{\mathrm{hel}}$ and $z_{\mathrm{CMB}}$ frames}{z hel and z CMB frames}}

\begin{figure*}
\centering

\begin{subfigure}{\textwidth}
    \centering
    \includegraphics[width=0.43\textwidth]{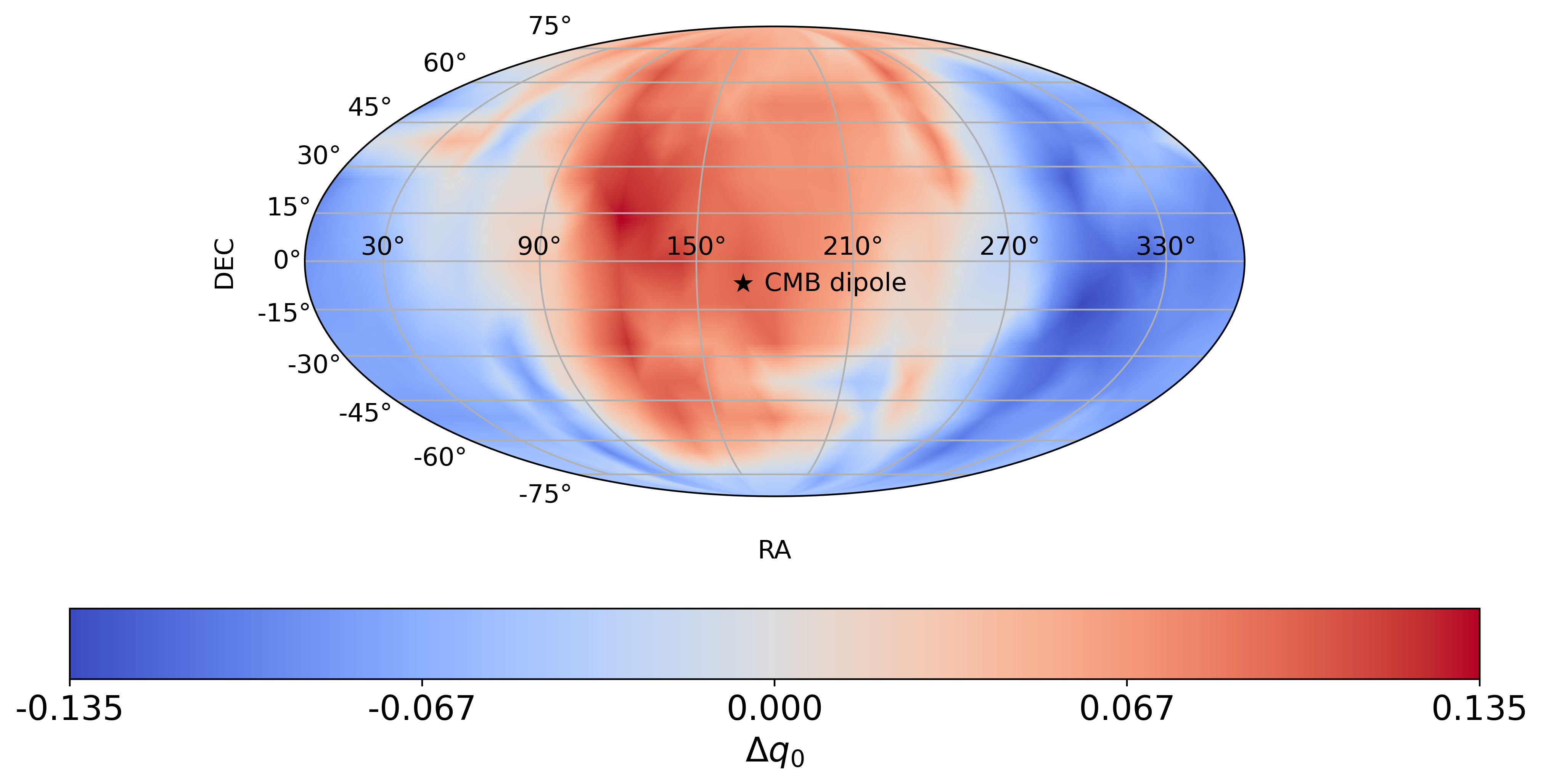}
    \hfill
    \includegraphics[width=0.43\textwidth]{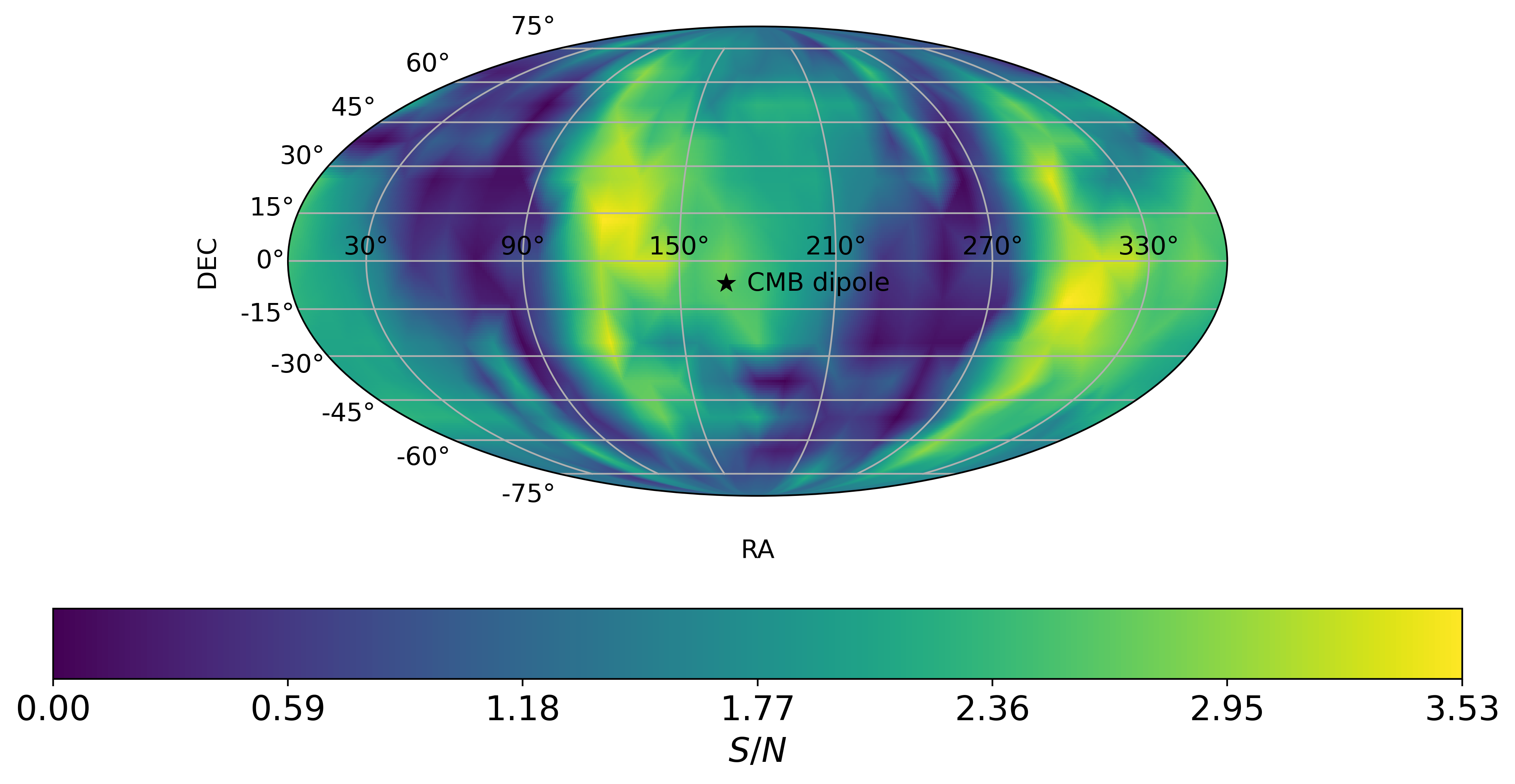}
    \caption{$0.01 < z_\textrm{hel} \leq 0.8$}
\end{subfigure}

\begin{subfigure}{\textwidth}
    \centering
    \includegraphics[width=0.43\textwidth]{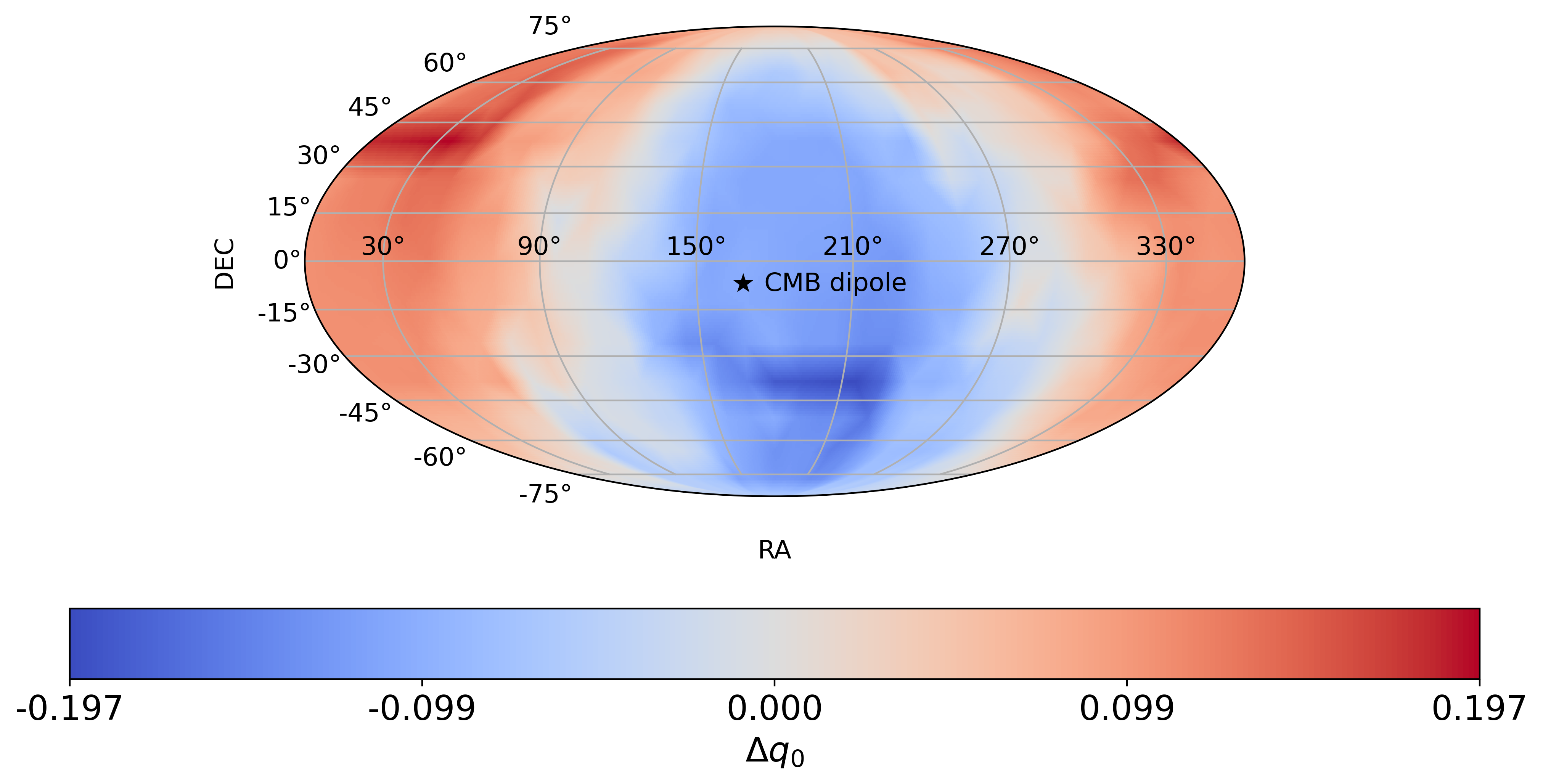}
    \hfill
    \includegraphics[width=0.43\textwidth]{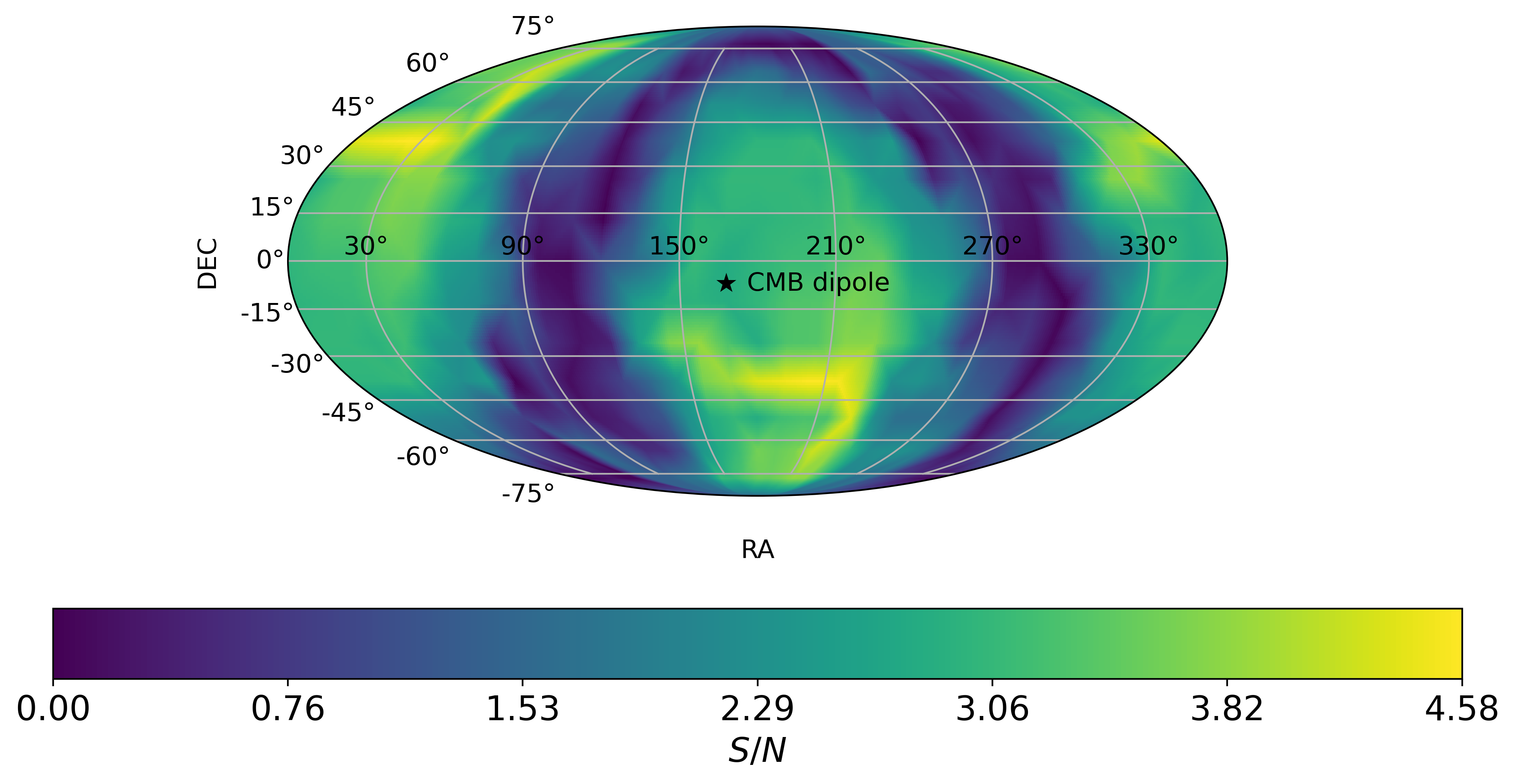}
    \caption{$0.01 < z_\textrm{CMB} \leq 0.8$}
\end{subfigure}
\caption{
Sky maps of the hemispherical contrast in $q_0$ for different redshift frames over the range $0.01 < z \leq 0.8$. Left column: $\Delta q_0 = q_0(\hat{\mathbf{n}}) - q_0(-\hat{\mathbf{n}})$. Right column: local signal to noise ratio $S/N$. Top row: using heliocentric redshifts $z_{\mathrm{hel}}$. Bottom row: using CMB frame redshifts $z_{\mathrm{CMB}}$. The star marks the direction of the CMB dipole. Compared to the $z_{\mathrm{HD}}$ frame, both $z_{\mathrm{hel}}$ and $z_{\mathrm{CMB}}$ exhibit stronger and more significant dipolar patterns, highlighting the role of kinematic effects and the impact of peculiar velocity corrections in suppressing hemispherical anisotropies.
}
\label{fig:deltaq0_sigma_zcmb_zhel}
\end{figure*}

We now extend the hemispherical analysis to alternative redshift frames in order to isolate the impact of kinematic corrections on the inferred anisotropy in $q_0$. In addition to the $z_{\mathrm{HD}}$ frame discussed above, we consider the heliocentric redshifts $z_{\mathrm{hel}}$ and the CMB frame redshifts $z_{\mathrm{CMB}}$. The latter are obtained by correcting $z_{\mathrm{hel}}$ for the observer’s motion inferred from the CMB dipole, using the velocity amplitude and direction measured by the \textit{Planck} collaboration \cite{Carr_redshifts,Planck_2018_I}. In the Pantheon+ compilation, this transformation defines the $z_{\mathrm{CMB}}$ frame. 

Since the full Pantheon+ covariance matrix, which includes both statistical and systematic contributions, incorporates terms associated with the modeling of peculiar velocities, we restrict, for this comparison, the analysis to the statistical covariance only, although we have verified that the results remain essentially unchanged when we use the full covariance matrix. This choice avoids introducing assumptions tied to a specific velocity field reconstruction when contrasting different redshift frames.

Fig.~\ref{fig:deltaq0_sigma_zcmb_zhel} presents the hemispherical contrast $\Delta q_0$ and the corresponding local signal to noise ratio $S/N$ for the frames $z_{\mathrm{hel}}$ and $z_{\mathrm{CMB}}$, both evaluated over the range $0.01 < z \leq 0.8$. In the heliocentric frame, a strong dipolar pattern is evident, with large extrema in both $\Delta q_0$ and $S/N$. Hemispheres aligned with the CMB dipole direction are associated with systematically lower values of $q_0$.

After transforming to the CMB frame, the dipolar structure remains pronounced and reaches an even larger statistical significance. In this case, hemispheres aligned with the CMB dipole exhibit higher values of $q_0$, indicating that correcting only for the observer's motion does not eliminate, and can even enhance, anisotropies induced by uncorrected peculiar velocities in the host galaxies.

Compared to these results, the corresponding maps in the $z_{\mathrm{HD}}$ frame show a clear reduction in both the amplitude and significance of the dipolar pattern. This progressive attenuation from $z_{\mathrm{hel}}$ to $z_{\mathrm{CMB}}$ and finally to $z_{\mathrm{HD}}$ provides direct evidence that a substantial fraction of the hemispherical anisotropy observed in $q_0$ arises from kinematic effects, and that correcting for peculiar velocities significantly mitigates these directional signatures.

\subsection{Inferred SNe dipole and its impact on hemispherical anisotropy \label{subsec:CMB_dipole_inference}}

We performed the inference of the SNe dipole, discussed in subsection \ref{subsec:SNe_dipole}, using an MCMC analysis in which the dipole parameters $(v_{\odot},\mathrm{RA}_{\rm dip}, \mathrm{DEC}_{\rm dip})$ are sampled simultaneously with the cosmographic parameters $q_0$ and $\mathcal{M}$. We adopt the cosmographic expansion for $m_\textrm{B}$, relation (\ref{eq:mB_cosmogra}), in the range $0.01 < z <0.8$, which is sufficient for the redshift range considered and avoids assumptions about the underlying cosmological model. We have explicitly verified that replacing the cosmographic description with a flat $\Lambda$CDM model, marginalizing over $H_0$ and $M_B$, leads to consistent constraints on the SNe dipole parameters, indicating that the inferred dipole is not driven by the specific background parametrization.

\begin{figure}[t]
    \centering
    \includegraphics[width=0.7\linewidth]{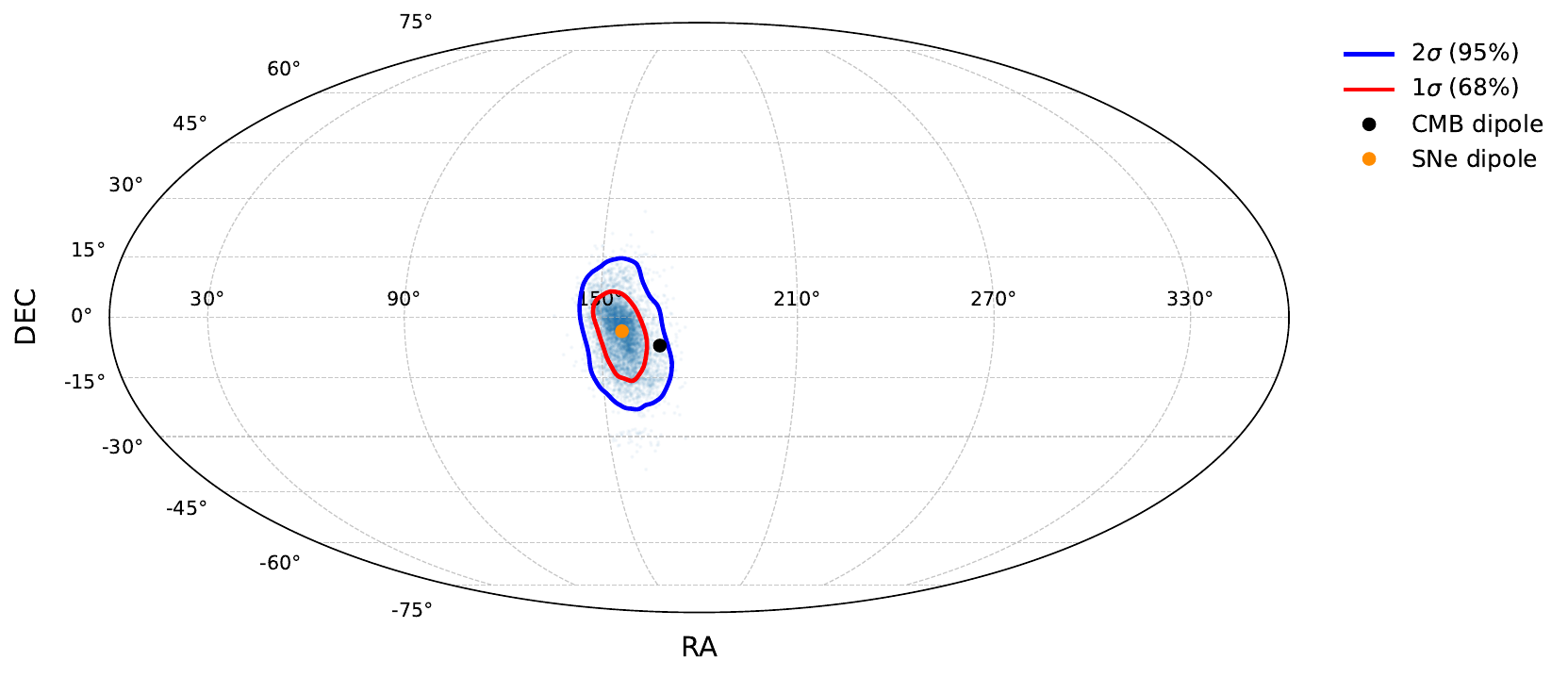}
    \caption{Sky map of the posterior constraints on the SNe dipole direction inferred from the MCMC analysis. The red and blue contours enclose the $1\sigma$ and $2\sigma$ confidence regions, respectively. The orange marker indicates the median SNe dipole direction from the posterior distribution, while the black marker shows the CMB dipole measured by Planck. The overlap between the contours and the Planck direction visually reflects the statistical consistency between both dipoles within current uncertainties, with the inferred dipole direction differing from the Planck value by $1.8\sigma$.}
    \label{fig:SNe_dipole_Mollweide_contours}
\end{figure}

From the MCMC posterior we obtain a median SNe dipole amplitude
\[
v_{\odot} = 307.26^{+32.00}_{-22.28}\,\mathrm{km\,s^{-1}},
\]
pointing towards
\[
(\mathrm{RA}_{\rm dip},\mathrm{DEC}_{\rm dip}) =
\bigl(156.40^{+4.72}_{-4.71},\, -3.38^{+5.54}_{-8.23}\bigr)^\circ,
\]
these values can be directly compared with the CMB dipole measured by Planck \cite{Planck_2018_I},
\begin{align*}
(v_{\odot})_{\rm Planck} &= 369.82 \pm 0.11\,\mathrm{km\,s^{-1}},\\
(\mathrm{RA}_{\rm Planck},\mathrm{DEC}_{\rm Planck})
&= (167.942^\circ \pm 0.011^\circ,\,-6.944^\circ \pm 0.005^\circ),
\end{align*}
we compute the difference between the minimum $\chi^2$ obtained when fitting the SNe dipole parameters and that obtained using the standard Pantheon+ $z_{\rm HD}$ redshifts based on the Planck dipole, finding $\Delta \chi^2 = \chi^2_{\rm SNe} - \chi^2_{\rm Planck} = -7.60,$
indicating that the SNe inferred dipole provides a better fit to the data. Interpreting this improvement in terms of a $\chi^2$ difference, this corresponds to a preference at the $\sim1.9\sigma$ level for the SNe dipole over the Planck dipole. A visual comparison of the directional constraints is shown in Figure \ref{fig:SNe_dipole_Mollweide_contours}, where the posterior contours of the SNe dipole at $1\sigma$ and $2\sigma$ confidence levels are displayed together with the CMB dipole direction.

Overall, both measurements are statistically consistent with the Planck CMB dipole. The moderately lower dipole amplitude inferred here is consistent with several SNe analyses. In particular, \cite{Horstmann_2022} finds a  similarly reduced dipole amplitude with a direction compatible with the CMB dipole, while \cite{Sorrenti_2023} reports amplitudes comparable to Planck but directions that depend sensitively on the adopted redshift cut and peculiar velocities treatment, differing at low redshift. By contrast, \cite{Singal_2022} finds a substantially larger dipole amplitude, of order $\sim 10^3\,\mathrm{km\,s^{-1}}$, albeit with a direction still broadly aligned with the CMB dipole. Taken together, these results indicate that while SNe based estimates of the dipole direction are generally consistent with the CMB within current uncertainties, the inferred amplitude exhibits significant scatter across analyses, likely reflecting sensitivity to low redshift effects and the  modeling of peculiar velocities.

Having inferred the SNe dipole amplitude and direction directly from the Pantheon+ sample,  we now use these values to construct an alternative set of Hubble diagram redshifts, $z_{\mathrm{HD}}^{\rm new}$, following the same relativistic and peculiar velocity correction pipeline described in subsection \ref{subsec:SNe_dipole}. In practice, the heliocentric to preferred SNe rest frame transformation and the subsequent peculiar velocity corrections are recomputed using the SNe dipole inferred values of $(v_{\odot}, \mathrm{RA}_{\mathrm{dip}}, \mathrm{DEC}_{\mathrm{dip}})$. Since the reconstructed peculiar velocity field is queried using the corrected redshift $z_{\mathrm{SN}}$ as input, changes in the dipole parameters also modify the inferred $v_p$ values. The resulting $z_{\mathrm{HD}}^{\rm new}$ therefore differs from the standard Pantheon+ $z_{\mathrm{HD}}$ both through the transformation to the observer frame and through the corresponding update of the peculiar velocity correction.

\begin{figure*}
\centering
\begin{subfigure}{\textwidth}
    \centering
    \includegraphics[width=0.43\textwidth]{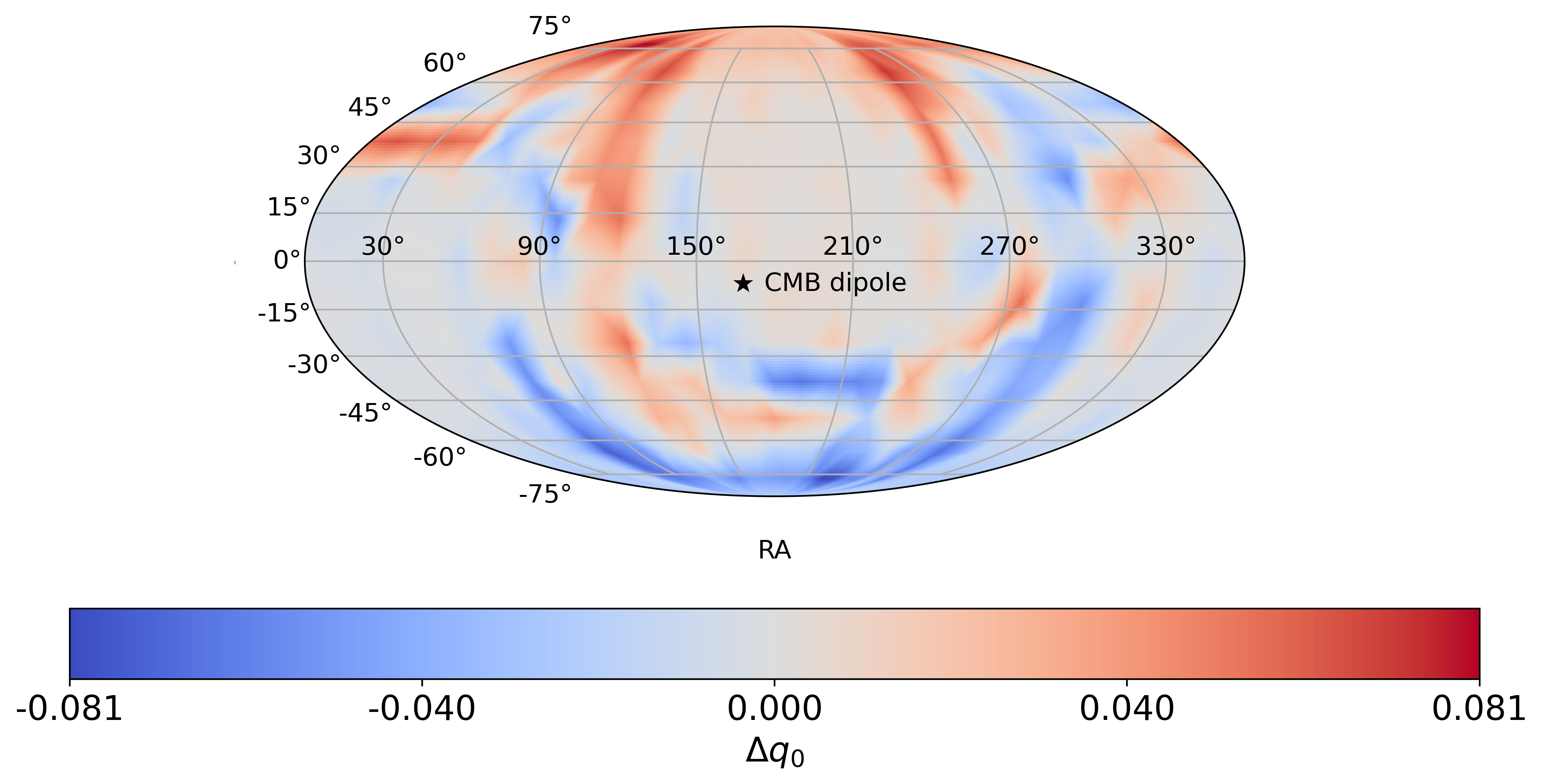}
    \hfill
    \includegraphics[width=0.43\textwidth]{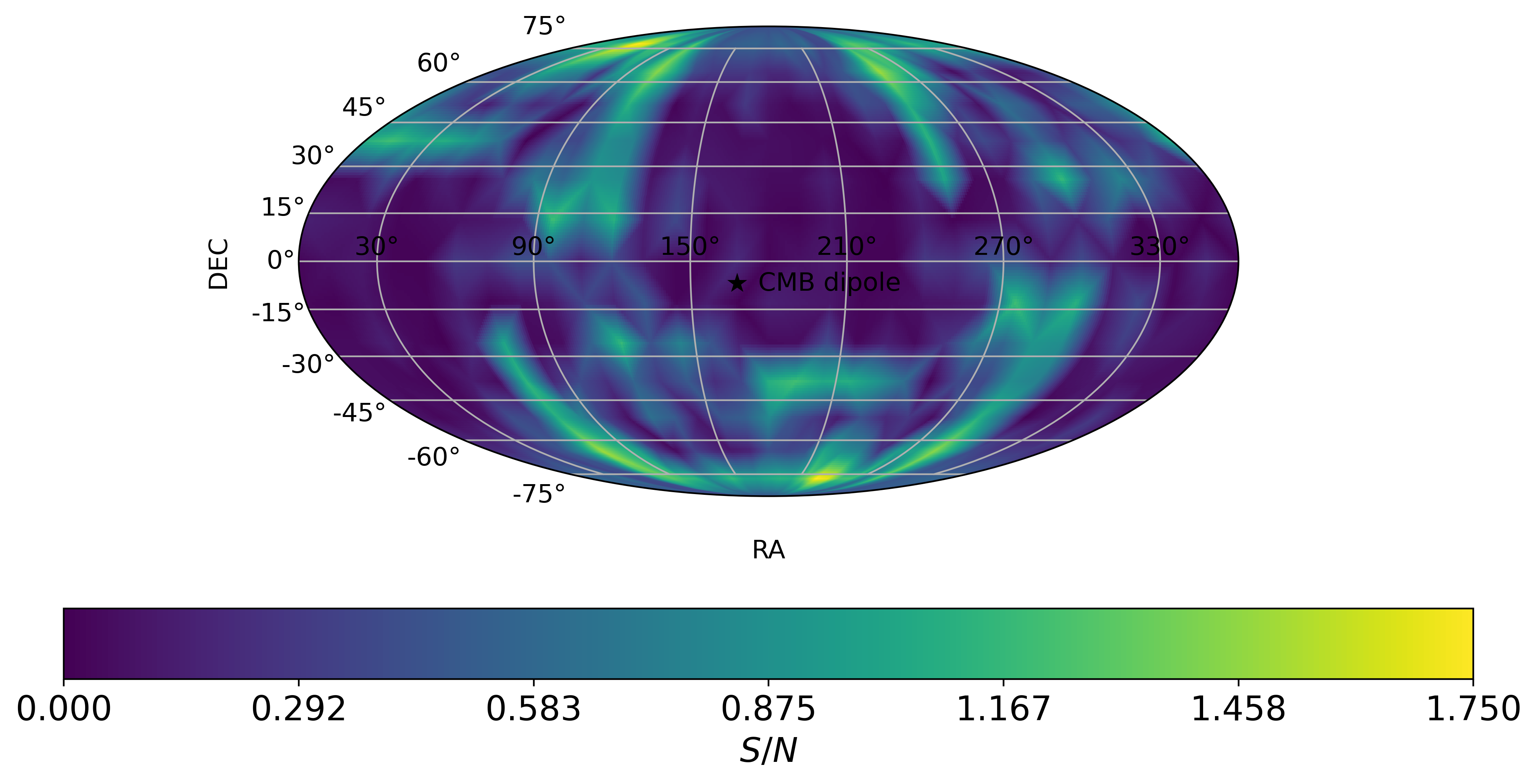}
\end{subfigure}
\caption{Sky maps of the hemispherical contrast in $q_0$ using the updated Hubble diagram redshifts $z_{\mathrm{HD}}^{\rm new}$ constructed with the SNe inferred CMB dipole parameters. Left column: $\Delta q_0 = q_0(\hat{\mathbf{n}}) - q_0(-\hat{\mathbf{n}})$. Right column: local signal to noise ratio $S/N$. The  dipolar pattern present in Fig.~\ref{fig:deltaq0_sigma_zcuts} is strongly suppressed, with the maximum local signal to noise ratio reduced to $|S/N| \lesssim 1.75$, indicating that the hemispherical signal is sensitive to the assumed CMB dipole parameters entering the redshift correction pipeline.
}
\label{fig:deltaq0_sigma_new_zHD}
\end{figure*}

We then repeat the hemispherical comparison analysis using these new redshifts, applying the same sky partition, fitting methodology, in the redshift cut $0.01 < z_{\mathrm{HD}}^{\rm new} < 0.8$. The resulting sky maps of $\Delta q_0$ and their associated local signal to noise ratio $S/N$ are shown in 
Fig.~\ref{fig:deltaq0_sigma_new_zHD}.

Compared to the analysis based on the standard Pantheon+ redshifts, the hemispherical pattern is markedly reduced. In particular, the large scale dipolar structure visible in the original $z_{\mathrm{HD}}$ maps is substantially suppressed, and the maximum local signal to noise ratio decreases to $|S/N| \lesssim 1.75$. The remaining fluctuations appear patchy and lack a coherent large scale alignment, consistent with expectations from statistical noise and residual sample variance.

This result suggests that a non negligible fraction of the previously observed hemispherical signal is sensitive to the assumed dipole parameters entering the redshift correction pipeline. In this sense, allowing the dipole amplitude and direction to vary within the SNe data themselves effectively absorbs part of the large scale anisotropy that would otherwise be interpreted as a dipolar modulation in $q_0$.

A natural interpretation is that the hemispherical signal detected in the standard Pantheon+ analysis does not necessarily reflect a genuine cosmological anisotropy, but may instead arise from subtle mismatches between the true local velocity field and the fixed dipole based corrections applied in the construction of $z_{\mathrm{HD}}$. In this sense, part of the apparent dipolar pattern could be absorbed or amplified depending on how the heliocentric to CMB transformation and the subsequent peculiar velocities corrections are implemented.

The fact that the SNe dipole amplitude is moderately lower than the value measured by Planck, albeit at only $\sim 2\sigma$ significance, highlights the sensitivity of SNe based redshift corrections to the modeling of local velocity fields. The Pantheon+ analysis emphasizes that the removal of kinematic dipole signatures through successive redshift corrections is not expected to be perfect, particularly at very low redshift where local structure and observational limitations can leave small residual directional patterns. Within this context, the partial suppression of the hemispherical signal when adopting the SNe inferred dipole parameters suggests that at least part of the anisotropic pattern observed in the standard Pantheon+ analysis may be linked to residual kinematic or local structure effects rather than to a genuine breakdown of large scale isotropy.

We emphasize that the present test does not demonstrate that the Pantheon+ redshift corrections are incorrect. Rather, it highlights that hemispherical anisotropy measurements are intrinsically sensitive to the details of the redshift correction procedure, and that even small variations in the assumed dipole parameters entering the pipeline can noticeably affect large scale anisotropy indicators. In this context, the disappearance of the dipolar pattern when using $z_{\mathrm{HD}}^{\rm new}$ suggests that at least part of the reported anisotropy may arise from residual kinematic effects associated with the redshift corrections. At the same time, the fact that the signal can be significantly altered by modifying the velocity field assumptions also leaves open the possibility that the observed hemispherical pattern could be tracing a more fundamental large scale effect, whose interpretation depends sensitively on the adopted reference frame and velocity modeling.

These findings motivate caution in interpreting dipolar signals in SNe datasets as evidence for departures from isotropy, and reinforce the importance of jointly assessing redshift corrections and anisotropy estimators within a unified and self consistent framework.

\subsection{Residual bulk flow from the SNe dipole constraints}

The results of the previous subsection show that the SNe inferred dipole is statistically consistent with, but not identical to, the standard CMB dipole. As discussed in Section~\ref{subsec:SNe_dipole}, such a mismatch can naturally arise if the effective frame selected by the Pantheon+ SNe sample contains residual anisotropic contributions not fully removed by the adopted peculiar velocity corrections, for example due to limitations in the reconstruction of the local velocity field or unmodeled coherent motions on large scales.

A simple phenomenological interpretation is to decompose the inferred SNe dipole as
\begin{equation}
\mathbf{D}_{\rm SNe}=\mathbf{D}_{\rm CMB}+\mathbf{D}_{\rm bulk},
\end{equation}
where $\mathbf{D}_{\rm CMB}$ is the standard kinematic dipole measured by Planck, while $\mathbf{D}_{\rm bulk}$ represents an additional residual contribution associated with coherent large scale motions or imperfections in the redshift correction pipeline. Using the median amplitudes and directions inferred above, the vector difference between the two dipoles gives
\begin{equation}
v_{\rm bulk}\simeq 94\,\mathrm{km\,s^{-1}},
\qquad
(\mathrm{RA}_{\rm bulk},\mathrm{DEC}_{\rm bulk})\simeq(31^\circ,16^\circ).
\end{equation}

To test this possibility directly, we introduce an additional coherent velocity component into the corrected Pantheon+ redshifts through 
\begin{equation}
1+z_{\rm HD}=\frac{(1+z_{\rm CMB})}{(1+z_p)(1+z_{\rm bulk})}\,,
\end{equation}
with
\begin{equation}
1+z_{\rm bulk}=\sqrt{\frac{1-\beta_{\rm bulk}}{1+\beta_{\rm bulk}}},
\qquad
\beta_{\rm bulk}=\frac{v_{\rm bulk}}{c}\,
\hat{\mathbf n}_{\rm SN}\!\cdot\!\hat{\mathbf n}_{\rm bulk},
\end{equation}
where $\hat{\mathbf n}_{\rm bulk}$ is specified by $(\mathrm{RA}_{\rm bulk},\mathrm{DEC}_{\rm bulk})$, $z_{\mathrm{CMB}}$ denotes the standard Pantheon+ redshifts in the CMB rest frame, constructed using a fixed dipole with amplitude and direction given by the Planck values, and $z_p$ represents the peculiar redshift induced by the line of sight peculiar velocity of the host galaxy, as discussed in Section~\ref{subsec:SNe_dipole}. We then infer $(v_{\rm bulk},\mathrm{RA}_{\rm bulk},\mathrm{DEC}_{\rm bulk})$ simultaneously with $(q_0,\mathcal M)$ using the same cosmographic likelihood, restricting the sample to $0.01<z_{\rm final}<0.8$.

Using an MCMC analysis, we obtain from the posterior distributions 
\begin{equation}
v_{\rm bulk}=98.78^{+35.66}_{-32.50}\,\mathrm{km\,s^{-1}},
\end{equation}
pointing towards
\begin{equation}
(\mathrm{RA}_{\rm bulk},\mathrm{DEC}_{\rm bulk})=
(29.37^{+22.75}_{-23.65},\,18.48^{+18.07}_{-26.25})^\circ .
\end{equation}

The excellent agreement between this direct fit and the independent estimate from $\mathbf{D}_{\rm SNe}-\mathbf{D}_{\rm CMB}$ strongly supports the interpretation that the small mismatch between the SNe and CMB dipoles can be understood as a residual bulk flow.

This amplitude, while small compared to the total Solar System motion inferred from the CMB dipole, differs from zero at the $\sim1.9\sigma$ level, providing evidence for a residual coherent velocity component not fully accounted for in the standard redshift correction pipeline.

Taken together, these results suggests that the difference between the SNe and CMB dipoles should not be interpreted as a negligible effect, but rather as a signature of residual systematics in the modeling of low redshift peculiar velocities. In this context, the inferred $\mathbf{D}_{\rm bulk}$ can be viewed as an effective correction that could be incorporated into the Pantheon+ analysis as an additional systematic component.

A complementary way to assess the impact of the reference frame choice is to compare the inferred values of the deceleration parameter $q_0$ obtained from the full Pantheon+ sample using different redshift frames. SH0ES reports $q_0 = -0.51 \pm 0.024$ when using the standard Pantheon+ redshifts $z_{\mathrm{HD}}$ \cite{Riess_2022}. Repeating the fit with alternative frames, and adopting the full statistical plus systematic covariance matrix, we obtain
\begin{eqnarray}
z_{\mathrm{hel}}:\; q_0 = -0.47 \pm 0.025, \\
z_{\mathrm{CMB}}:\; q_0 = -0.48 \pm 0.024, \\
z_{\mathrm{HD}}:\; q_0 = -0.51 \pm 0.024, \\
z_{\mathrm{HD}}^{\mathrm{new}}:\; q_0 = -0.50 \pm 0.024.
\end{eqnarray}

These results show that the inferred value of $q_0$ is sensitive at the $\Delta q_0 \sim 0.01$--$0.04$ level to the adopted redshift frame. In particular, using the SNe inferred dipole to construct $z_{\mathrm{HD}}^{\mathrm{new}}$ shifts the result by $\Delta q_0 \simeq +0.01$ relative to the standard SH0ES value, partially restoring consistency with the baseline analysis.

As we mentioned above, this variation can be interpreted as a residual systematic effect associated with the modeling of low redshift velocity corrections. In the Pantheon+ analysis \cite{Pantheon+}, peculiar velocity uncertainties are incorporated into the covariance matrix through an effective dispersion term $\sigma_{\rm vpec} \simeq 240 \, \mathrm{km \,s^{-1}}$ as well as limited systematic variations obtained by comparing alternative density map reconstruction such as the 2MRS \cite{2MRS_densisty_map}. However, the dipole used to define the CMB frame is kept fixed, and the velocity field corrections rely on a specific model of the local density distribution. Our results suggest that residual large scale coherent motions or mismatches between the true velocity field and its reconstruction, such as incomplete sky coverage or unmodeled bulk flows, may not be fully captured by this treatment. Although subdominant compared to statistical uncertainties, this residual effect is non negligible and highlights the sensitivity of anisotropy measurements to the assumptions entering the redshift correction pipeline.

\section{\label{sec:conclusion}Summary and conclusions}

In this work, we have investigated the isotropy of the present day deceleration parameter, $q_0$, using a hemispherical comparison method applied to the Pantheon+ SNe Ia compilation. Motivated by recent reports of large scale anisotropies in cosmological observables and by theoretical expectations from tilted cosmological models \cite{Tsagas_2010, Tsagas_2011, Tsagas_2015, Tsagas_2021}, our analysis focused on identifying possible dipolar signatures in $q_0$ and assessing their robustness under different redshift frames.

We implemented a directional hemispherical scan over the full sky, fitting a cosmographic expansion of the apparent magnitude $m_\textrm{B}$ independently in opposite hemispheres and constructing sky maps of the hemispherical contrast $\Delta q_0$ and its local signal to noise ratio $S/N$. Special attention was paid to the role of redshift corrections, comparing results obtained in the heliocentric frame $z_{\mathrm{hel}}$, CMB frame $z_{\mathrm{CMB}}$, and fully corrected Hubble diagram redshifts $z_{\mathrm{HD}}$, which include corrections for the peculiar velocities of the host galaxies, including the impact of progressively excluding low redshift SNe, where peculiar velocities effects are expected to be most relevant.

When using the standard Pantheon+ $z_{\mathrm{HD}}$ redshifts, we find evidence for a dipole like hemispherical pattern in $q_0$ at low minimum redshift cuts, Fig. \ref{fig:deltaq0_sigma_zcuts}, where the direction of maximum asymmetry broadly aligned with the CMB dipole. The amplitude and statistical significance of this pattern decreases as more conservative redshift cuts are imposed, and the signal becomes indistinguishable from statistical fluctuations for $z_{\mathrm{min}} \gtrsim 0.05$. This behavior is consistent with an origin driven primarily by local or intermediate scale effects, rather than by a genuinely large scale violation of isotropy.

By explicitly comparing different redshift frames, we showed that the hemispherical aniso\-tropy is strongest in the heliocentric and CMB frame redshifts, see Fig. \ref{fig:deltaq0_sigma_zcmb_zhel} and is substantially reduced once corrections for host galaxy peculiar velocities are applied. This demonstrates that a significant fraction of the observed directional signal can be attributed to kinematic effects and highlights the critical role of peculiar velocity modeling in anisotropy studies based on low and intermediate redshift SNe.

It is worth noting that the removal of the kinematic dipole signal through the heliocentric to CMB and subsequent peculiar velocity corrections is not expected to be perfect at the lowest redshifts. As discussed in subsection 4.6 of \cite{Pantheon+}, Pantheon+ finds that although the dominant dipolar pattern present in heliocentric redshifts is largely suppressed after applying the corrections to $z_{\mathrm{CMB}}$ and $z_{\mathrm{HD}}$, a small residual signal remains visible in the direction opposite to the CMB dipole. The residual signal is highly local, confined to $z \lesssim 0.02$ and increasing toward lower redshifts, with the largest deviations around $z \simeq 0.01$. Possible explanations include incomplete modeling of the coupling between the Milky Way motion and the motions of nearby galaxies, low number statistics (the reported deviation is at the $\sim 1\sigma$ level), and uneven sky coverage of the SNe sample, particularly due to clustering in the Stripe 82 region. Importantly, these residuals are driven almost entirely by SNe at $z < 0.02$, which are excluded from the SH0ES sample and therefore do not affect the local $H_0$ inference. These considerations indicate that, although standard redshift corrections significantly suppress kinematic dipole signatures, small residual anisotropies may persist at very low redshift due to astrophysical and observational limitations, motivating further investigations of dipolar patterns using controlled redshift cuts as pursued in this work.

As an internal consistency test of the Pantheon+ redshift correction framework, we allowed the dipole that best maps the observed heliocentric redshifts into an isotropic Hubble diagram frame, which we denote as the \emph{SNe dipole}, to be inferred directly from the SNe data rather than fixed to the Planck CMB values. This quantity need not coincide exactly with the standard CMB dipole, since in addition to the Solar System motion it may absorb residual anisotropic contributions associated with imperfect peculiar velocity corrections or unresolved local large scale motions. For this inference we adopted the full statistical plus systematic covariance matrix provided by Pantheon+, ensuring consistency with the baseline likelihood and allowing a direct internal comparison with the published analysis. This isolates the effect of varying the dipole parameters while leaving the rest of the pipeline unchanged. A fully self consistent treatment would require recomputing the covariance matrix for each dipole realization, since changes in the heliocentric transformation and predicted peculiar velocities can propagate into the velocity dependent covariance terms. Such a recalculation is computationally beyond the scope of the present work. From the MCMC posterior we obtain a median SNe dipole amplitude $v_{\odot}=307.26^{+32.00}_{-22.28}\,\mathrm{km\,s^{-1}},$ pointing toward $(\mathrm{RA}_{\rm dip},\mathrm{DEC}_{\rm dip})= \bigl(156.40^{+4.72}_{-4.71},\, -3.38^{+5.54}_{-8.23}\bigr)^\circ$. These values can be directly compared with the Planck CMB dipole measurement,$(v_{\odot})_{\rm Planck}=369.82\pm0.11\,\mathrm{km\,s^{-1}}$ and $ (\mathrm{RA}_{\rm Planck},\mathrm{DEC}_{\rm Planck})=(167.94^\circ,\,-6.94^\circ)$, see Fig. \ref{fig:SNe_dipole_Mollweide_contours}. We find that the SNe dipole direction differs from the Planck direction at approximately $1.8\sigma$, while the amplitude shows a slightly larger deviation of about $\sim2\sigma$. Overall, the two determinations remain statistically compatible within current uncertainties, but the mild offset suggests that the effective frame preferred by the SNe data is not necessarily identical to the standard CMB frame.

The moderately lower dipole amplitude obtained here is broadly consistent with several recent SNe based analyses \cite{Sorrenti_2023, Horstmann_2022}, although the literature shows significant scatter in the inferred amplitudes, likely reflecting the sensitivity of dipole measurements to low redshift effects and to the modeling of peculiar velocities in both the redshift corrections and the covariance matrix. By contrast, analyses based on large samples of quasars and radio galaxies typically report significantly larger dipole amplitudes than expected from a purely kinematic interpretation of the CMB dipole, often at the level of several times the Planck amplitude, albeit with directions broadly aligned with it \cite{Singal_2011, Bengaly_2018, Singal_PhysRevD.100.063501, Schwarz_refId0, Secrest_2021, Singa_10.1093/mnras/stac144, 10.1111/j.1365-2966.2012.22032.x, Secrest_2022, Colin_10.1093/mnras/stx1631,Tiwari_2016}.

Using these inferred SNe dipole parameters to reconstruct an alternative set of Hubble diagram redshifts, $z_\mathrm{HD}^{\mathrm{new}}$, produces a striking suppression of the hemispherical anisotropy signal. As shown in Figure~\ref{fig:deltaq0_sigma_new_zHD}, the prominent large scale dipolar pattern present in the original maps is strongly reduced, while the maximum local signal to noise ratio decreases to $|S/N|\lesssim 1.75$. The residual structure is patchy and lacks a coherent preferred direction, consistent with fluctuations expected from statistical noise and sample variance. This demonstrates that a substantial fraction of the previously reported anisotropy is sensitive to the dipole assumptions entering the redshift correction pipeline. More specifically, the suppression may arise from an imperfect match between the fixed observer dipole adopted from Planck and the effective frame preferred by the SNe sample, from residual inaccuracies in the reconstructed peculiar velocity field, or from the coupling between both ingredients in the low redshift corrections. 

A natural step is to test whether the mild offset between the SNe and CMB dipoles can be interpreted as a modest residual coherent velocity component. To do so, we adopt the phenomenological decomposition $\mathbf{D}_{\rm SNe}=\mathbf{D}_{\rm CMB}+\mathbf{D}_{\rm bulk},$ where $\mathbf{D}_{\rm bulk}$ represents an additional residual contribution beyond the standard CMB dipole correction. The vector difference between the two dipoles then gives an independent estimate $v_{\rm bulk}\simeq 94\,\mathrm{km\,s^{-1}}$ pointing towards $(\mathrm{RA}_{\rm bulk}, \mathrm{DEC}_{\rm bulk})\simeq(31^\circ,16^\circ),$ fully consistent with the direct MCMC analysis including an additional bulk flow contribution in the redshift corrections, $ v_{\rm bulk}=98.78^{+35.66}_{-32.50}\,\mathrm{km\,s^{-1}}, (\mathrm{RA}_{\rm bulk},\mathrm{DEC}_{\rm bulk})= (29.37^{+22.75}_{-23.65},\,18.48^{+18.07}_{-26.25})^\circ.$ This concordance strongly supports the view that the small mismatch between the SNe and CMB dipoles can be explained by residual low redshift velocity systematics, such as imperfect peculiar velocity corrections or unresolved coherent motions. At the same time, the fact that the hemispherical signal changes so significantly under these physically motivated reanalyses leaves open the possibility that part of the original pattern could also be tracing a genuine large scale anisotropy, for example associated with deviations from statistical isotropy, anisotropic cosmic expansion as predicted in tilted cosmological scenarios or other models with preferred directions, residual inhomogeneities or large scale inhomogeneities beyond the standard $\Lambda$CDM framework, whose inferred significance depends sensitively on the adopted reference frame and velocity modeling.

A complementary test of the robustness of the inferred deceleration parameter is provided by comparing the value of $q_0$ obtained from the full Pantheon+ sample under different redshift constructions. In particular, comparing the standard Pantheon+ redshifts $z_{\mathrm{HD}}$ with those reconstructed using the SNe inferred dipole, $z_{\mathrm{HD}}^{\mathrm{new}}$, we find a shift of $\Delta q_0 \simeq 0.01$. This comparison isolates the impact of the dipole assumptions within the same correction framework, indicating that part of the inferred value of $q_0$ is sensitive to the specific choice of dipole parameters entering the redshift pipeline. For completeness, we also find larger variations, $\Delta q_0 \sim 0.01$–$0.04$, when using alternative frames such as $z_{\mathrm{hel}}$ and $z_{\mathrm{CMB}}$, as expected given their incomplete kinematic corrections. Although the shift between the standard $z_{\mathrm{HD}}$ and $z_{\mathrm{HD}}^{\mathrm{new}}$ is smaller than the statistical uncertainty, it reflects a coherent effect and highlights the sensitivity of SNe based cosmological inferences to the details of the redshift correction procedure.

Taken together, our findings suggest that the hemispherical anisotropy detected in $q_0$ using standard Pantheon+ redshifts does not provide robust evidence for a fundamental breakdown of large scale isotropy. Instead, it is likely driven, at least in part, by residual kinematic effects and by the detailed implementation of redshift and peculiar velocity corrections. This interpretation is consistent with expectations from tilted cosmological models \cite{Tsagas_2010, Tsagas_2011, Tsagas_2015, Tsagas_2021}, in which relative motions between observers can induce apparent dipolar modulations in locally inferred cosmological parameters without requiring departures from an underlying FLRW geometry. In this context, part of the observed anisotropic signal could also be associated with large scale inhomogeneities in the matter distribution, which can generate effective anisotropies in low redshift observables such as SNe. While our results highlight the dominant role of reference frame choices and velocity systematics, they do not fully exclude the possibility that a fraction of the signal reflects a genuine large scale effect.

More broadly, this work highlights the strong coupling between anisotropy estimators and the choice of observational frame in SNe based analyses. Using hemispherical comparisons, we find a dipolar anisotropy in $q_0$ reaching $\Delta q_0 = 0.112$ with a maximum significance of $S/N \simeq 2.16$ in the standard Pantheon+ $z_{\mathrm{HD}}$ frame, aligned with the CMB dipole direction and decreasing as low redshift SNe are removed. This signal is substantially reduced when adopting a dipole inferred directly from the SNe data or, equivalently, when allowing for a residual bulk velocity component of order $\sim 100 \, \mathrm{km \,s^{-1}}$, indicating that a significant fraction of the observed anisotropy can be attributed to kinematic effects associated with the redshift correction pipeline. Hemispherical tests of cosmological parameters, while powerful, must therefore be interpreted with caution and within a fully self consistent treatment of redshift corrections. Future SNe samples, with improved sky coverage, higher redshift reach, and better control of peculiar velocity systematics, such as those expected from upcoming wide field surveys, will be essential for placing more stringent and robust constraints on directional variations in the cosmic expansion and for disentangling kinematic effects from potential contributions due to large scale structure.

\bibliographystyle{JHEP}
\bibliography{bib.bib}

\end{document}